\documentclass{article}
\usepackage[ascii]{inputenc}
\usepackage[T1]{fontenc}
\usepackage[french,ngerman,italian,english]{babel}
\usepackage{amsmath,amssymb,amsfonts,textcomp}
\usepackage{color}
\usepackage{array}
\usepackage{supertabular}
\usepackage{hhline}
\usepackage{hyperref}
\usepackage[]{graphicx}
\newcommand\textstyleInternetlink[1]{\textcolor{blue}{#1}}
\newcommand\textstyleHTMLZitat[1]{\textit{#1}}
\makeatletter
\newcommand\arraybslash{\let\\\@arraycr}
\makeatother
\newcounter{saveenum}
\newcommand\liststyleWWviiiNumii{%
\renewcommand\theenumi{\arabic{enumi}}
\renewcommand\theenumii{\alph{enumii}}
\renewcommand\theenumiii{\roman{enumiii}}
\renewcommand\theenumiv{\arabic{enumiv}}
\renewcommand\labelenumi{\theenumi.}
\renewcommand\labelenumii{\theenumii.}
\renewcommand\labelenumiii{\theenumiii.}
\renewcommand\labelenumiv{\theenumiv.}
}
\makeatletter
\newcommand\ps@Standard{
  \renewcommand\@oddhead{}
  \renewcommand\@evenhead{}
  \renewcommand\@oddfoot{\hfill \hfill {}- \thepage{}}
  \renewcommand\@evenfoot{\@oddfoot}
  \renewcommand\thepage{\arabic{page}}
}
\makeatother
\newcommand\normalsubformula[1]{\text{\mathversion{normal}$#1$}}
\title{Phylogenetic Applications of the Minimum Contradiction Approach on Continuous Characters}

\begin{document}
\clearpage\setcounter{page}{1}\pagestyle{Standard}

\maketitle

\begin{center} 
{\small{\selectlanguage{english}
To appear in \href{http://www.la-press.com/evolutionary-bioinformatics-journal-j17}{\textstyleInternetlink{\foreignlanguage{english}{\textit{Evolutionary Bioinformatics}}}}  2009}}
\end{center}

\bigskip

\bigskip

{\selectlanguage{ngerman}
Marc Thuillard}

{\selectlanguage{english}
\foreignlanguage{ngerman}{La Colline, 2072
St-Blaise}\foreignlanguage{ngerman}{ (Switzerland)
}\href{mailto:Thuillweb@hotmail.com}{\textstyleInternetlink{\foreignlanguage{ngerman}{Thuillweb@hotmail.com}}}}

\bigskip

{\selectlanguage{french}
Didier Fraix-Burnet}

{\selectlanguage{english}
Universit\'e Joseph Fourier, CNRS, Laboratoire
d{\textquotesingle}Astrophysique de Grenoble, BP53, F-38041 Grenoble
(France) }\href{mailto:fraix@obs.ujf-grenoble.fr}{\textstyleInternetlink{\foreignlanguage{english}{fraix@obs.ujf-grenoble.fr}}}

\bigskip

{\selectlanguage{english}
\textbf{Abstract:} We describe the conditions under which a set of
continuous variables or characters \ can be described as an X-tree or
\ a split network. A distance matrix corresponds exactly to a split
network or a valued X-tree if, after ordering of the taxa, the
variables values can be embedded into a function with at most a local
maxima and a local minima, and crossing any horizontal line at most
twice. In real applications, the order of the taxa best satisfying the
above conditions can be obtained using the Minimum Contradiction
method. This approach is applied to 2 sets of continuous characters.
The first set corresponds to craniofacial landmarks in Hominids. The
contradiction matrix is used to identify possible tree structures and
some alternatives when they exist. We explain how to discover the main
structuring characters in a tree. The second set consists of a sample
of 100 galaxies. In that second example one shows how to discretize the
continuous variables describing physical properties of the galaxies
without disrupting the underlying tree structure. \ }

\bigskip

{\selectlanguage{english}
\foreignlanguage{english}{\textbf{1}}\foreignlanguage{english}{\textbf{.
Introduction}}}

{\selectlanguage{english}
Maximum parsimony and distance-based approaches are the most popular
methods to produce phylogenetic trees. Whereas most studies use
discrete characters, there is a growing need for applying phylogenetic
methods to continuous characters. Examples of continuous data include
gene expressions (Planet et al. 2001), gene frequencies (Edwards and
Cavalli-Sforza 1964; 1967), phenotypic characters (Oakley and
Cunningham, 2000) or some morphologic characters (MacLeod and Forey
\ 2003; Gonz\'alez-Jos\'e et al. 2008).\foreignlanguage{english}{ }}

{\selectlanguage{english}
The simplest method to deal with continuous characters using maximal
parsimony consists of discretizing the characters into a number of
states small enough to be processed by the software. Recent software
programs such as TNT (Tree analysis using New Technology; Goloboff et
al. 2008) or CoMET (Continuous-character Model Evaluation and Testing
Model; Lee and al. 2007) \foreignlanguage{english}{use developments of
the contrast method to deal with continuous characters. These methods
assume that the characters evolve at comparable rates according to a
Brownian motion, an assumption that is often difficult to verify
(Felsenstein, 2004; Oakley \ and Cunningham, 2000). Distance-based
methods are applied to both discrete and continuous input data.
Compared to character-based approaches, distance-based approaches are
quite fast and furnish in many instances quite reasonable results. As
pointed out by Felsenstein (2004), the amount of information that is
lost when using a distance-based algorithm compared to a
character-based approach is often surprisingly small. The use of
continuous characters in distance-based methods may at first glance be
less problematic than in character-based methods, since algorithms like
the Neighbour-Joining work identically on discrete or continuous
characters. However, here too it is often not easy to determine if the
data can be described by a tree. When does a set of continuous
\ characters describe a split network or an
}\foreignlanguage{english}{X-tree? \ The article furnishes some new
insights on that question. It explains when a set of continuous
characters can be described exactly by a split network or a valued
X-tree. In real applications, the distance matrix corresponds only
approximately to a split network or a tree topology. An adequate method
is necessary to quantify to what extent the distance matrix corresponds
to a split network or a tree. The Minimum Contradiction method can be
used for that purpose (Thuillard, 2007; 2008; 2009). }}

{\selectlanguage{english}
\foreignlanguage{english}{The}\foreignlanguage{english}{ paper is
organized as follows. Section 2 succinctly presents the Minimum
Contradiction method. It explains why some inequalities, called
Kalmanson inequalities, \ are central to phylogenies. Section 3 extends
the Minimum Contradiction method to a set of continuous characters.
Section 4 furnishes the conditions under which a set of continuous
characters can be described by a tree or a phylogenetic network.
Section 5 presents an application of the algorithms in morphometrics
using a set of faciocranial characters of hominids. Section 6 presents
preliminary results on the evolution of a number of physical characters
in galaxies. It illustrates how the Minimum Contradiction approach can
be applied to discover structuring characters.}}

\bigskip

{\selectlanguage{english}
\foreignlanguage{english}{\textbf{2}}\foreignlanguage{english}{\textbf{.
Ordering the \ taxa on a tree or a split network}}}

{\selectlanguage{english}
\foreignlanguage{english}{A valued \ X}\foreignlanguage{english}{{}-tree
T is a graph with X the set of leaves and a unique path between any two
distinct vertices x and y, with internal vertices of at most degree 3.
A circular order on an X-tree corresponds to an indexing of the n
leaves according to a circular (clockwise or anti-clockwise) scanning
of the leaves in T (Makarenkov and Leclerc, 1997). Figure 1 shows a
tree and an indexing of the taxa that corresponds to a circular order.
For taxa indexed according to a circular order the distance matrix }
${Y_{{\normalsubformula{\text{i,j}}}}^{{n}}}$\foreignlanguage{english}{
fulfils the so-called Kalmanson inequalities (Kalmanson, 1975):}}

{\selectlanguage{english}
 ${Y_{{i,j}}^{{n}}\overset{}{{}}\ge
\overset{}{{}}Y_{{i,k}}^{{n}}}$\foreignlanguage{english}{, }
${Y_{{k,j}}^{{n}}\overset{}{{}}\ge
\overset{}{{}}Y_{{k,i}}^{{n}}}$\foreignlanguage{english}{ (} ${i\le
j\le k}$\foreignlanguage{english}{) with } ${Y_{{i,j}}^{{n}}=1/2\cdot
(d_{{i,n}}+d_{{j,n}}-d_{{i,j}})}$\foreignlanguage{english}{.  
\ \ (1)}}

{\selectlanguage{english}
\foreignlanguage{english}{with } ${d_{{i,j}}}$
\foreignlanguage{english}{the pairwise distance between taxon i and j.
As depicted in Fig.1, the matrix element }
${Y_{{\normalsubformula{\text{i,j}}}}^{{n}}}$\foreignlanguage{english}{
is the distance between a reference node n and the path i-j. The
diagonal elements }
${Y_{{i,i}}^{{n}}=d_{{i,n}}}$\foreignlanguage{english}{ correspond to
the pairwise distance between the reference node and the taxon i. The
distance matrix }
${Y_{{\normalsubformula{\text{i,j}}}}^{{n}}}$\foreignlanguage{english}{
has the property that the distance diminishes away from the diagonal
(Kalmanson, 1975). This property is visualized in Fig 1. If the values
of the distance matrix are \ represented by different levels of gray,
the level of gray is shading away from the diagonal. This property of
the matrix characterizes a Kalmanson matrix and an order satisfying all
Kalmanson inequalities is called a perfect order. }}

\bigskip

{\selectlanguage{english}
\includegraphics[width=9.433cm,height=5.103cm]{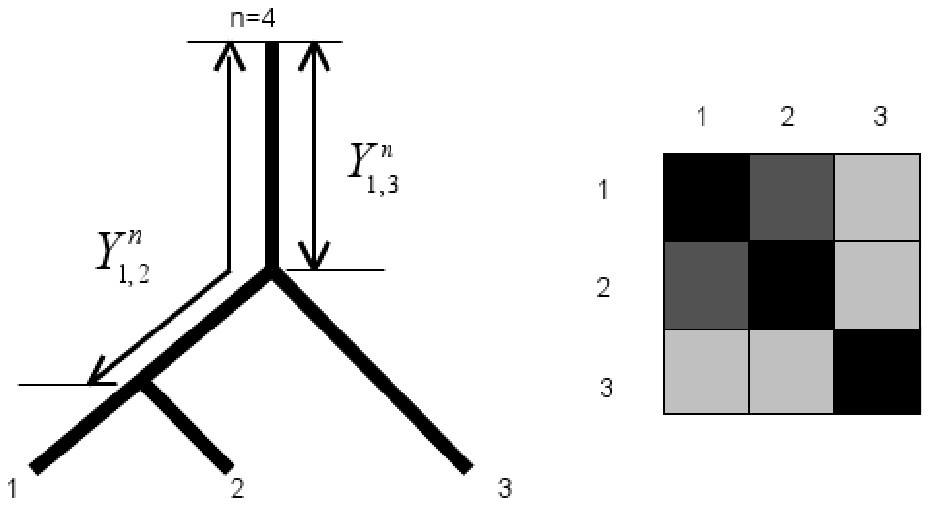}
 }

\bigskip

{\selectlanguage{english}
\foreignlanguage{english}{Figure 1.}\foreignlanguage{english}{ The
distance }
${Y_{{\normalsubformula{\text{i,j}}}}^{{n=4}}}$\foreignlanguage{english}{
between a reference taxa n and the path i-j on an X-tree fulfils
Kalmanson inequalities. If the values of the distance matrix }
${Y_{{\normalsubformula{\text{i,j}}}}^{{n=4}}}$
\foreignlanguage{english}{are coded in a gray scale, the level of gray
decreases as one moves away from the diagonal. For more details see
Thuillard (2007).}}

\bigskip

{\selectlanguage{english}
\foreignlanguage{english}{In real applications, the \ distance matrix }
${Y_{{\normalsubformula{\text{i,j}}}}^{{n}}}$
\foreignlanguage{english}{often only partially fulfils the inequalities
corresponding to a perfect order. The contradiction on the order of the
taxa can be defined \ as}}

\includegraphics[width=6cm]{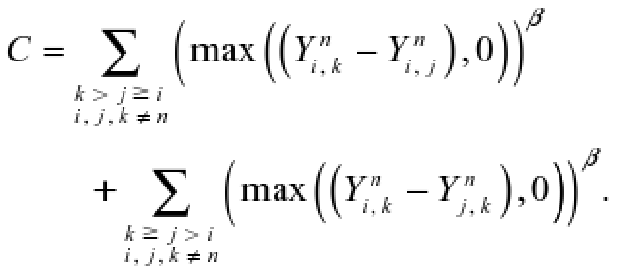}     \ \ \ (2)

\bigskip

{\selectlanguage{english}
\foreignlanguage{english}{The best order of a distance matrix is, by
definition, the order minimizing the contradiction. The ordered matrix
}
${Y_{{\normalsubformula{\text{i,j}}}}^{{n}}}$\foreignlanguage{english}{
corresponding to the best order is defined as the minimum contradiction
matrix for the reference taxon n. For a perfectly ordered X-tree, the
contradiction C is zero. A high contradiction value C is the indication
of a distance matrix deviating significantly from an X-tree. Bandelt
and Dress (1992) have shown that if a distance matrix }
${d_{{\normalsubformula{\text{i,j}}}}}$\foreignlanguage{english}{
fulfils Kalmanson inequalities, then the distance matrix can be exactly
represented by a split network or by an X-tree. A split network can be
regarded as a generalization of trees. A split is a partition of the
taxa into two disjoint sets that is realized by removing the edges
relating the two sets. (For an introduction to split networks, see
Huson and Bryant, 2006). Kalmanson inequalities are related to a number
of interesting mathematical results. Kalmanson inequalities relate
phylogenetic trees and split networks to the travelling salesman
problem. Let us recall that the travelling salesman problem is a
fundamental problem in computer science. The problem{\textquoteright}s
formulation is quite simple. A travelling salesman must visit a number
of cities and return to its point of departure. The problem consists of
finding the order of the cities that minimizes the total travelling
distance }
${D=d_{{n,1}}+\underset{{i=1,\text{.}\text{.}\text{.},(n-1)}}{\sum
}{d_{{i,i+1}}}}$\foreignlanguage{english}{ with }
${d_{{i,j}}}$\foreignlanguage{english}{ the distance between the city i
and j. The travelling salesman is one of the most studied problem in
computational science as it is the prototype of a difficult problem.
\ For all known algorithms, the maximum computing time to solve the
travelling salesman problem increases very rapidly with the number of
cities. In other words, the solution of the travelling salesman problem
for a large number of cities generally requires a very large computing
power. Already for a few hundreds cities, only approximate solutions
can be obtained by the largest computers. Not all TSP problems are
difficult to solve. For instance, the TSP is easy to solve when the
cities are on a convex hull in the Euclidean plane. In order to be on a
convex hull, the cities must be orderable so that the following
inequalities hold: } ${d_{{i,j}}+d_{{k,n}}\le
d_{{i,k}}+d_{{j,n}}}$\foreignlanguage{english}{ and }
${d_{{i,n}}+d_{{j,k}}\le d_{{i,j}}+d_{{k,n}}}$
\foreignlanguage{english}{with } ${1\le i\le j\le k\le
n}$\foreignlanguage{english}{ (Kalmanson,
}\foreignlanguage{english}{1975). These inequalities are equivalent to
the Kalmanson inequalities (1): } ${Y_{{i,j}}^{{n}}\ge
\overset{}{{}}Y_{{i,k}}^{{n}}}$\foreignlanguage{english}{; \ }
${Y_{{k,j}}^{{n}}\ge
\overset{}{{}}Y_{{k,i}}^{{n}}}$\foreignlanguage{english}{ ( \ } ${i\le
j\le k\le n}$\foreignlanguage{english}{). The solution to the TSP
corresponds to the order of the cities on the convex hull.}}

\bigskip

\includegraphics[width=8cm]{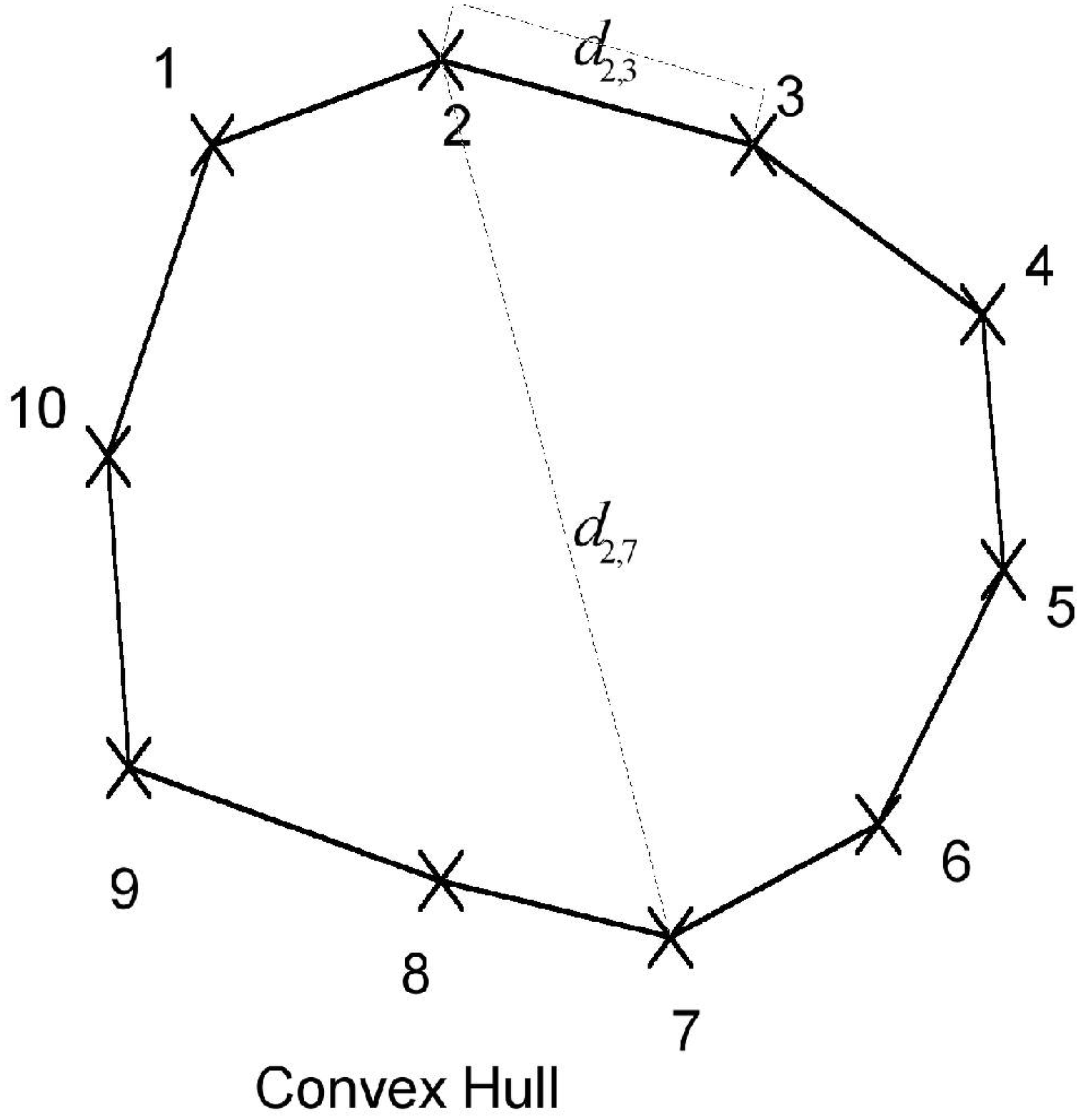}

\bigskip

{\selectlanguage{english}
\foreignlanguage{english}{Figure 2.}\foreignlanguage{english}{ The
travelling salesman problem (TSP) can be easily solved if the points
are on a convex hull in the Euclidean plane. Points on a convex hull
fulfil the Kalmanson inequalities.}}

\bigskip

{\selectlanguage{english}
\foreignlanguage{english}{If one leaves aside Euclidian geometry, other
metrics fulfil Kalmanson inequalities. Kalmanson inequalities are also
satisfied by taxa on an X-tree or a split network. If the taxa are
circularly ordered, then the Kalmanson inequalities are fulfilled. As
developed in a number of publications (Deineko et al. 1995; Christopher
et al.1996; Dress and Huson, 2004), perfect order corresponds in
X-trees \ and split networks to a solution of the travelling salesman
problem (TSP) for both the distance matrices }
${d_{{\normalsubformula{\text{i,j}}}}}$\foreignlanguage{english}{ and }
${Y_{{\normalsubformula{\text{i,j}}}}^{{n}}}$\foreignlanguage{english}{.
\ }}

{\selectlanguage{english}
\foreignlanguage{english}{In the next section we show that for trees and
split networks as well, the Kalmanson inequalities are related to
convexity. This result furnishes a new perspective on when trees and
phylogenetic networks can be used to describe a set of continuous
characters. \ }}

\bigskip

{\selectlanguage{english}
\foreignlanguage{english}{\textbf{3}}\foreignlanguage{english}{\textbf{.
Kalmanson inequalities on \ a single continuous character}}}

{\selectlanguage{english}
\foreignlanguage{english}{As of today, i}\foreignlanguage{english}{t is
still not really clear when the use of continuous characters in
distance-based phylogenetic studies is a valid approach. To clarify
that problem, we will first consider a single character. }}

{\selectlanguage{english}
\foreignlanguage{english}{Let us now discuss the conditions for which a
set of taxa characterized by a single continuous character }
${f_{{1}}}$\foreignlanguage{english}{ can be perfectly ordered. Let us
define \ the distance } ${d_{{i,j}}}$\foreignlanguage{english}{ between
two taxa as }
${d_{{i,j}}=\normalsubformula{\text{abs}}(f(i)-f(j))}$\foreignlanguage{english}{.
The taxa \{1,..,n\} are perfectly ordered when the order is such that
the distance matrix } ${Y_{{i,j}}^{{n}}}$\foreignlanguage{english}{
fulfils the Kalmanson inequalities: } ${Y_{{i,j}}^{{n}}\ge
\overset{}{{}}Y_{{i,k}}^{{n}}}$\foreignlanguage{english}{, }
${Y_{{k,j}}^{{n}}\overset{}{{}}\ge
\overset{}{{}}Y_{{k,i}}^{{n}}}$\foreignlanguage{english}{
\ \ \ \ \ \ \ \ \ \ \ (} ${i\le j\le k\le
n}$\foreignlanguage{english}{). Proposition 1 describes the necessary
and sufficient conditions on the character} ${f_{{1}}(i)}$
\foreignlanguage{english}{so that the taxa can be perfectly ordered.}}

\bigskip

{\selectlanguage{english}
Proposition 1:}

{\selectlanguage{english}
\foreignlanguage{english}{A distance matrix }
${Y_{{i,j}}^{{n}}}$\foreignlanguage{english}{ is Kalmanson if and only
if the values } ${f_{{1}}(i)}$ \foreignlanguage{english}{of a character
on an ordered set of taxa can be embedded into a continuous function }

${f(x)}$\foreignlanguage{english}{ on [1,n]: } ${f(x)=(x-i)\cdot
(f(i+1)-f(i))+f(i),\overset{}{{}}x\in [i,i+1],\overset{}{{}}x\subset
\Re ,i\in \{1,\text{.}\text{.}\text{.},n\}}$\foreignlanguage{english}
{
with the following properties:}}

\bigskip

{\selectlanguage{english}
\foreignlanguage{english}{i) the function } ${f(x)}$
\foreignlanguage{english}{has \ at most one local maxima and one local
minima}}

{\selectlanguage{english}
\foreignlanguage{english}{ii) the function} ${f(x)}$
\foreignlanguage{english}{crosses the reference line }
${L(x)=f_{{1}}(n)=\normalsubformula{\text{const}}\text{.}}$
\foreignlanguage{english}{at most once.}}

\bigskip

{\selectlanguage{english}
\foreignlanguage{english}{Proof}\foreignlanguage{english}{\textbf{:}}}

{\selectlanguage{english}
\foreignlanguage{english}{A central distinction can be made between the
taxa depending}\foreignlanguage{english}{ on whether the character
value is smaller or larger than the value of a reference taxon n. The
set of taxa can be divided into two disjoint sets, the set S of taxa
with values smaller or equal to the reference value }
${f_{{1}}(n)}$\foreignlanguage{english}{ \ and the set of \ taxa L with
values larger than the reference value (See Fig. 3 for an
illustration). Let us show that a distance matrix fulfilling the
conditions i) and ii) is perfectly ordered for any 3 ordered taxa }
${i\le j\le k}$\foreignlanguage{english}{. We will consider all
possible cases}}

{\selectlanguage{english}
\foreignlanguage{english}{a) All 3 taxa are in the same set (S or L).
The distance } ${Y_{{i,j}}^{{n}}}$ \foreignlanguage{english}{between
the taxa i and j \ \ is given by the expression }
${Y_{{i,j}}^{{n}}=\text{min}(|f_{{1}}(i)-f_{{1}}(n)|,|f_{{1}}(j)-f_{{1}}(n)|)}$\foreignlanguage{english}{.
Under the conditions in Prop.1 one has }
${\text{min}(|f_{{1}}(i)-f_{{1}}(n)|,|f_{{1}}(j)-f_{{1}}(n)|)\ge
\text{min}(|f_{{1}}(i)-f_{{1}}n)|,|f_{{1}}(k)-f_{{1}}(n)|)}$}

{\selectlanguage{english}
\foreignlanguage{english}{and }\foreignlanguage{english}{consequently}
${Y_{{i,j}}^{{n}}\ge
\overset{}{{}}Y_{{i,k}}^{{n}}}$\foreignlanguage{english}{, ( \ } ${i\le
j\le k\le n}$\foreignlanguage{english}{). }}

{\selectlanguage{english}
\foreignlanguage{english}{b) The taxon}\foreignlanguage{english}{ i is
in one set of taxa and the taxa j,k in another set. In that case one
has \ }
${Y_{{i,j}}^{{n}}=Y_{{i,k}}^{{n}}=0\text{.}}$\foreignlanguage{english}{
(For an illustration, see Fig.5 and Eq.3) }}

{\selectlanguage{english}
c) Condition ii) prevents the second taxon to be in another set than the
taxa i and k. }

{\selectlanguage{english}
\foreignlanguage{english}{d) If the third taxa is in another set than
the taxa i,j one has } ${Y_{{i,j}}^{{n}}\ge
\overset{}{{}}Y_{{i,k}}^{{n}}=0}$\foreignlanguage{english}{. The proof
for the second inequality } ${Y_{{k,j}}^{{n}}\overset{}{{}}\ge
\overset{}{{}}Y_{{k,i}}^{{n}}}$\foreignlanguage{english}{ ( \ } ${i\le
j\le k\le n}$\foreignlanguage{english}{) is similar. }}

{\selectlanguage{english}
\foreignlanguage{english}{Let us show that if the conditions of the
proposition are not fulfilled then Kalmanson inequalities are violated.
If the function } ${f(x)}$\foreignlanguage{english}{ has two maxima (or
2 minima) corresponding to the taxa i and k, then there exists a taxa j
with } ${Y_{{i,j}}^{{n}}<Y_{{i,k}}^{{n}}}$\foreignlanguage{english}{
\ and consequently the Kalmanson inequalities are not fulfilled. A
similar inequality holds if the function } ${f(x)}$
\foreignlanguage{english}{does not satisfy condition ii).}}

\bigskip

\bigskip

{\selectlanguage{english}
\foreignlanguage{english}{Figure 3 illustrates
Prop.}\foreignlanguage{english}{ 1 with a simple example. The matrix }
${Y_{{i,j}}^{{n}}}$ \foreignlanguage{english}{is depicted using a
colour coding. Large values are coded red, while small values of }
${Y_{{i,j}}^{{n}}}$\foreignlanguage{english}{ correspond to small
values. The distance matrix is perfectly ordered; the values of }
${Y_{{i,j}}^{{n}}}$\foreignlanguage{english}{ decrease away from the
diagonal as prescribed by the Kalmanson inequalities. Two clusters are
observed, the first cluster corresponds to values smaller than the
reference value, the second cluster to values larger than the reference
value. \ }}

{\selectlanguage{english}
\foreignlanguage{english}{The results on a single character can be
easily generalized to several characters as the sum of perfectly
ordered matrices }
${Y_{{i,j}}^{{n}}=\overset{{m\text{max}}}{\underset{{m=1}}{\sum
}}{Y_{{i,j}}^{{n}}(f_{{m}})}}$ \foreignlanguage{english}{is also
perfectly ordered. This follows directly from the Kalmanson
inequalities. If each character is Kalmanson, then }
${Y_{{i,j}}^{{n}}(f_{{m}})\ge \overset{}{{}}Y_{{i,k}}^{{n}}(f_{{m}})}$
\foreignlanguage{english}{and }
${Y_{{k,j}}^{{n}}(f_{{m}}\overset{}{{{)}}}\ge
\overset{}{{}}Y_{{k,i}}^{{n}}(f_{{m}})}$\foreignlanguage{english}{ (
\ } ${i\le j\le k\le n}$\foreignlanguage{english}{), and therefore }
${Y_{{i,j}}^{{n}}}$\foreignlanguage{english}{ is perfectly ordered. }}

\bigskip

 \includegraphics[width=10cm]{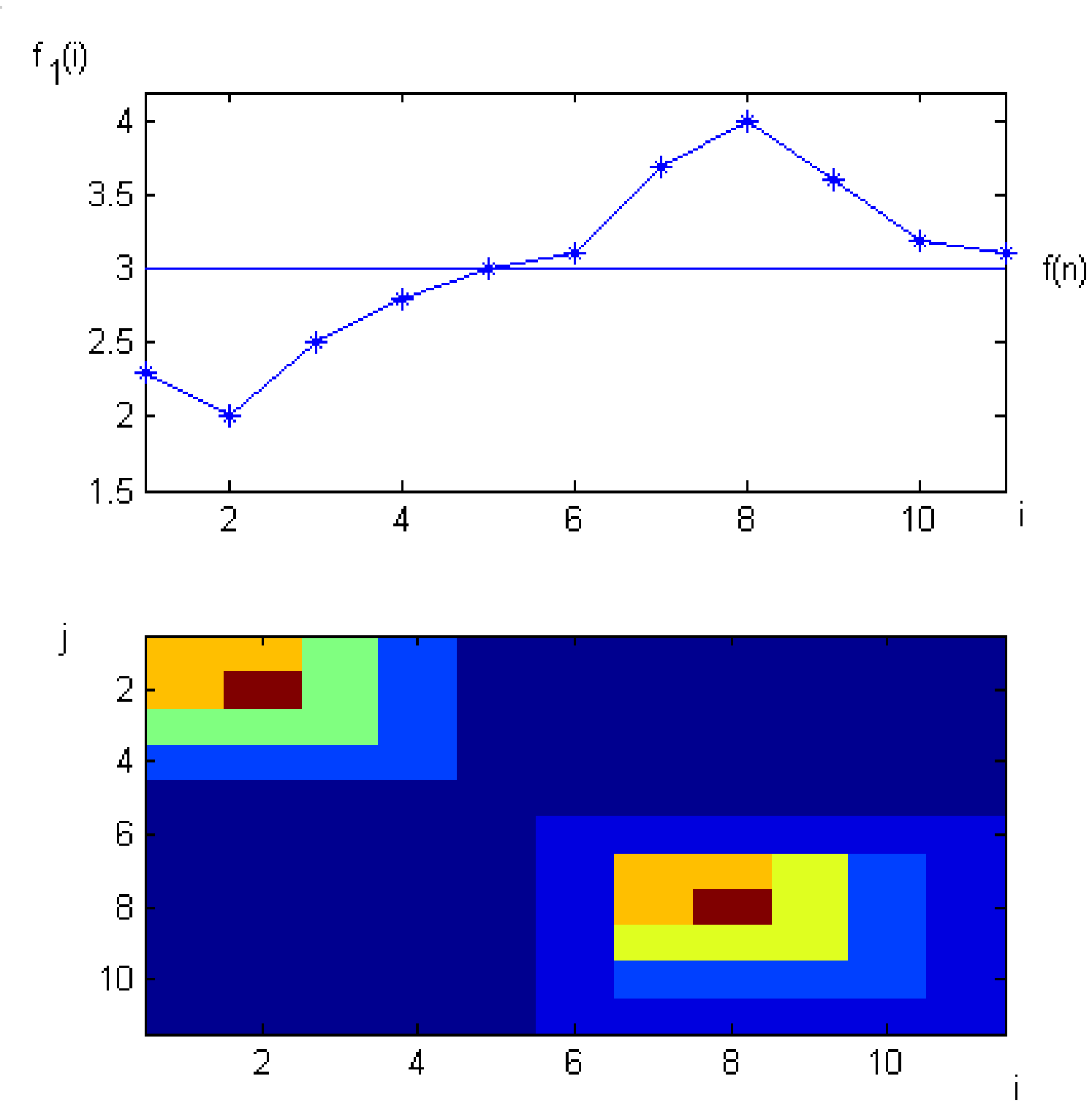} 
\includegraphics[width=0.441cm,height=4.482cm]{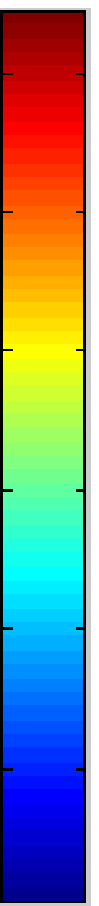}

{\selectlanguage{english}
\foreignlanguage{english}{Figure 3.}\foreignlanguage{english}{ Top: The
taxa are ordered so that the characters } ${f_{{1}}(i)}$
\foreignlanguage{english}{on the taxa \{1,{\dots},i,{\dots},n\} can be
embedded in a function } ${f(x)}$ \foreignlanguage{english}{fulfilling
proposition 1. Bottom: Distance matrix } ${Y_{{i,j}}^{{n}}}$
\foreignlanguage{english}{with a colour coding. Larger values are coded
red, small values blue. The order is perfect (C=0 in Eq.2).}}

\bigskip

{\selectlanguage{english}
\foreignlanguage{english}{We are now ready to discuss the connection
between Kalmanson inequalities an}\foreignlanguage{english}{d convexity
in phylogenies. The tree metrics case \ is different from the Euclidean
metrics described in Fig.2. In an Euclidean metrics, Kalmanson
inequalities are fulfilled if the points (cities) are on a convex hull,
while for split networks and trees the hull must be orthogonally
convex. In an Euclidean metrics, a set } ${Z\subset \Re ^{{n}}}$
\foreignlanguage{english}{is defined to be orthogonally convex if, for
every line that is parallel to one of the axes of the Cartesian
coordinate system, the intersection of }
${Z}$\foreignlanguage{english}{ with the line \ is empty, a point, or a
single interval. }}

\bigskip

\bigskip

{\selectlanguage{english}
Corollary 2:}

{\selectlanguage{english}
\foreignlanguage{english}{If the taxa \{1,{\dots},n\} are ordered so
that the distance matrices }
${Y_{{i,j}}^{{n}}}$\foreignlanguage{english}{ associated to the 2
characters } ${f_{{1}}}$\foreignlanguage{english}{ and }
${f_{{2}}}$\foreignlanguage{english}{ are perfectly ordered, then the
closed circuit }
${\{(f_{{1}}(1),f_{{2}}(1));\text{.}\text{.}\text{.};(f_{{1}}(n),f_{{2}}(n)\}}$\foreignlanguage{english}{
relating each two \ consecutive points \ by an edge is on an orthogonal
convex hull. }}

\bigskip

{\selectlanguage{english}
Proof: }

{\selectlanguage{english}
\foreignlanguage{english}{Proposition 1 for a single character is
equivalent to the following proposition: if the distance matrix }
${Y_{{i,j}}^{{n}}}$\foreignlanguage{english}{ associated to a character
} ${f_{{1}}}$\foreignlanguage{english}{ is Kalmanson, then any
horizontal line crosses the function } ${f(x)}$
\foreignlanguage{english}{at most once (see Fig. 3 for an
illustration). It follows that any horizontal or vertical line in the
Euclidian plane intersects the closed curve }
${\{(f_{{1}}(1),f_{{2}}(1));\text{.}\text{.}\text{.};(f_{{1}}(n),f_{{2}}(n)\}}$\foreignlanguage{english}{
at most twice. (The intersection of the line with }
${Z}$\foreignlanguage{english}{ is either a single interval or a point
or empty (no crossing)). Let us point out that Corollary 2 describes a
sufficient but not necessary condition to obtain a perfectly ordered
matrix } ${Y_{{i,j}}^{{n}}}$\foreignlanguage{english}{.}}

\includegraphics[width=12cm]{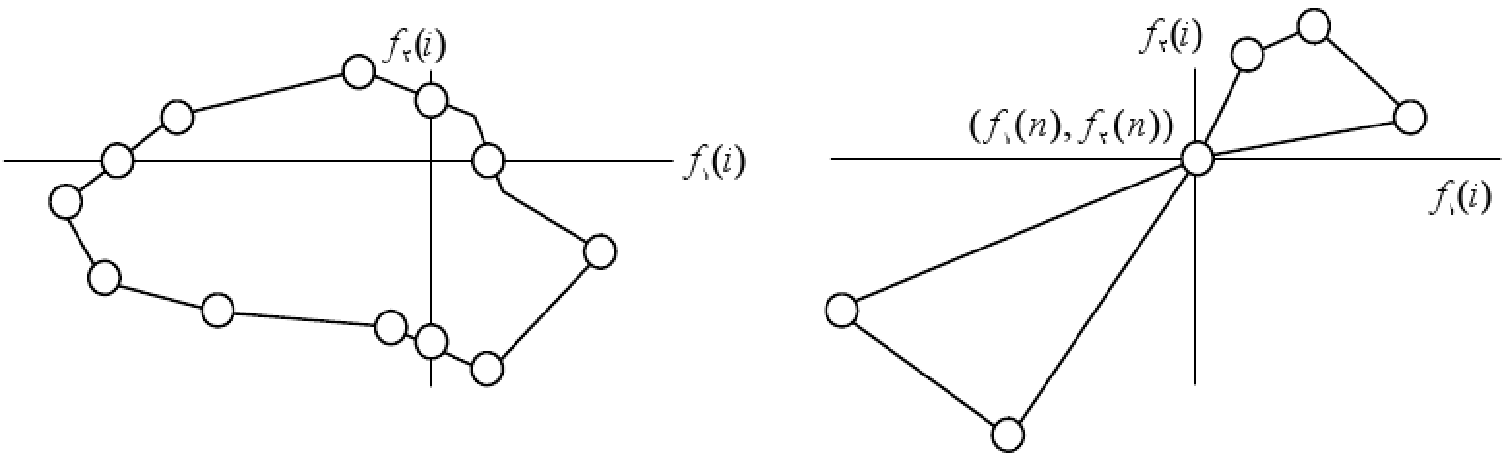}

\bigskip

{\selectlanguage{english}
\foreignlanguage{english}{Figure 4.}\foreignlanguage{english}{ The
values of two characters that are perfectly ordered are on an
orthogonal convex hull. Two examples of an orthogonal convex hulls.}}

\bigskip

{\selectlanguage{english}
\foreignlanguage{english}{Corollary 2 can be extended to higher
dimensions. The geometry, associated to trees and split networks built
on a set of perfectly ordered characters, corresponds to an
orthogonally convex hull. }}

\bigskip

{\selectlanguage{english}
\foreignlanguage{english}{\textbf{4}}\foreignlanguage{english}{\textbf{.
How to build a tree or a phylogenetic network from single continuous
characters? \ }}}

{\selectlanguage{english}
\foreignlanguage{english}{In the previous section we have explained when
a}\foreignlanguage{english}{ set of characters on a set of taxa
\ fulfils Kalmanson inequalities and can be described by a \ tree or a
split network. In this section, we explicitly show how the branches of
the trees evolve when several characters are combined. For a single
character, the taxa can be ordered so as to fulfil the conditions of
Prop. 1. The resulting tree is a line tree. In a line tree, all \ taxa
are on a single path and one has}}

{\selectlanguage{english}
 ${Y_{{i,j}}^{{n}}=\begin{matrix}0\overset{}{{}}i\in S,j\notin
S\overset{}{{}}\normalsubformula{\text{or}}\overset{}{{}}i\in L,j\notin
L\\\text{min}(|f(i)-f(n)|,|f(j)-f(n)|)=\text{min}(Y_{{i,i}}^{{n}},Y_{{j,j}}^{{n}})\overset{}{{}}\normalsubformula{\text{otherwise}}\end{matrix}}$.
  \ \ \ (3)}

{\selectlanguage{english}
Figure 5 shows an example \ of a line tree with perfectly ordered taxa.
}

\bigskip

\includegraphics[width=12cm]{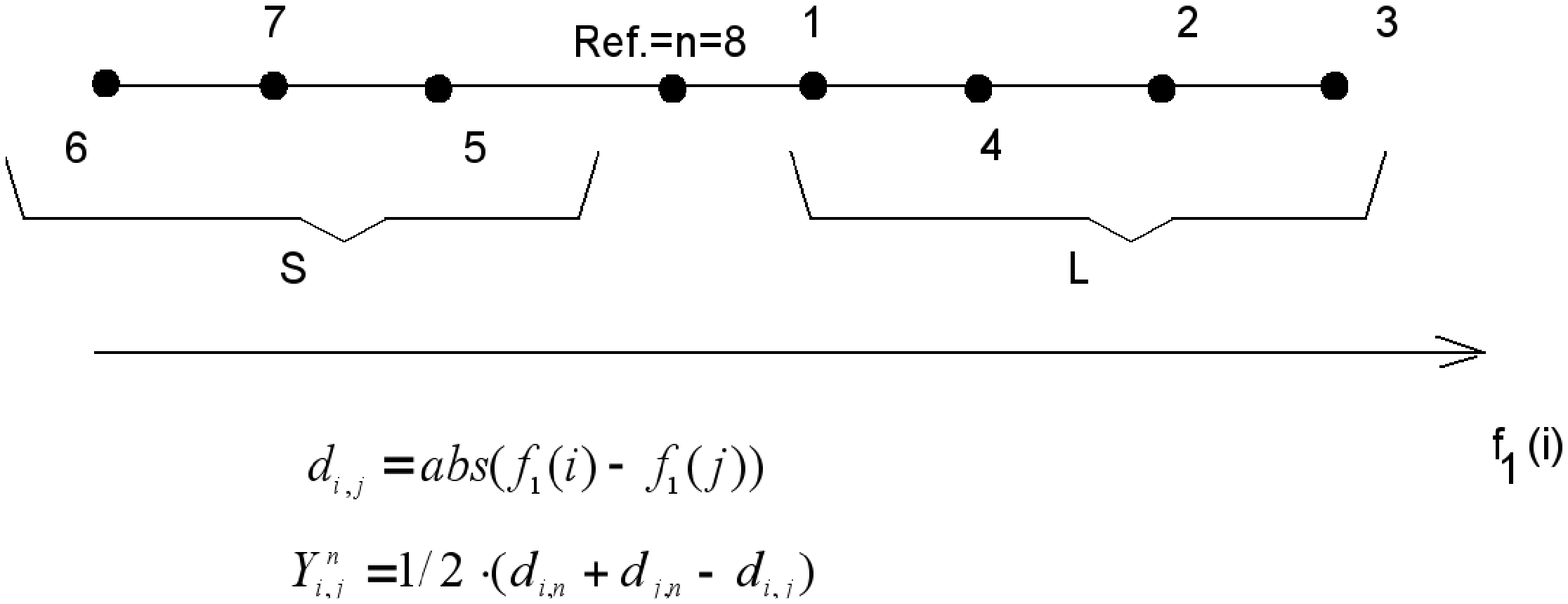}

{\selectlanguage{english}
\foreignlanguage{english}{Figure 5.}\foreignlanguage{english}{ The tree
associated to a single character is a line tree. \ In a line tree, all
taxa are on the same path. }}

\bigskip

{\selectlanguage{english}
At least two independent characters are necessary to generate a tree
that is not a line tree. An independent character can be defined as
follows. }

{\selectlanguage{english}
\foreignlanguage{english}{Definition}\foreignlanguage{english}{
1:}\foreignlanguage{english}{ }}

{\selectlanguage{english}
\foreignlanguage{english}{Two characters }
${f_{{1}}}$\foreignlanguage{english}{ and }
${f_{{2}}}$\foreignlanguage{english}{ are independent if there exists
at least 2 taxa i and j }}

{\selectlanguage{english}
\foreignlanguage{english}{(i{\textless}j{\textless}n) so that }
${0<Y_{{i,j}}^{{n}}<Y_{{i,i}}^{{n}},Y_{{j,j}}^{{n}}}$\foreignlanguage{english}{
with }
${Y_{{i,j}}^{{n}}=Y_{{i,j}}^{{n}}(f_{{1}})+Y_{{i,j}}^{{n}}(f_{{2}})}$\foreignlanguage{english}{.}}

{\selectlanguage{english}
Proposition 3:}

{\selectlanguage{english}
\foreignlanguage{english}{If \ t}\foreignlanguage{english}{wo
characters} ${f_{{1}}}$\foreignlanguage{english}{ and } ${f_{{2}}}$
\foreignlanguage{english}{are independent, then the distance matrix }
${Y_{{i,j}}^{{n}}=Y_{{i,j}}^{{n}}(f_{{1}})+Y_{{i,j}}^{{n}}(f_{{2}})}$\foreignlanguage{english}{
does not correspond to a line tree.}}

\bigskip

{\selectlanguage{english}
Proof:}

{\selectlanguage{english}
\foreignlanguage{english}{A line tree }\foreignlanguage{english}{is so
that either} ${Y_{{i,j}}^{{n}}=0}$\foreignlanguage{english}{ or }
${Y_{{i,j}}^{{n}}=\text{min}(Y_{{i,i}}^{{n}},Y_{{j,j}}^{{n}})}$\foreignlanguage{english}{.
By definition two independent characters do not fulfil either
equality.}}

\bigskip

{\selectlanguage{english}
\foreignlanguage{english}{Figure 6a shows 3 examples of independent
}\foreignlanguage{english}{characters. If two characters are
independent and the taxa are perfectly ordered on both }
${f_{{1}}}$\foreignlanguage{english}{ and }
${f_{{2}}}$\foreignlanguage{english}{, then the distance matrix
corresponds to a split network or an X-tree different from a line tree.
\ Let us discuss the first example in Fig.6. Without restriction, let
us assume that for the reference taxon n, }
${f_{{1}}(n)=f_{{2}}(n)=0}$\foreignlanguage{english}{. The distance
matrix elements are given by}}

{\selectlanguage{english}

${Y_{{i,j}}^{{n}}=\left(\begin{matrix}f_{{1}}(i)+f_{{2}}(i)&\text{min}(f_{{1}}(i),f_{{1}}(j))+\text{min}(f_{{2}}(i),f_{{2}}(j))\\\text{min}(f_{{1}}(i),f_{{1}}(j))+\text{min}(f_{{2}}(i),f_{{2}}(j))&f_{{1}}(j)+f_{{2}}(j)\end{matrix}\right)}$\foreignlanguage{english}{.
The expression reduces to }
${Y_{{i,j}}^{{n}}=\left(\begin{matrix}f_{{1}}(i)+f_{{2}}(i)&f_{{1}}(j)+f_{{2}}(i)\\f_{{1}}(j)+f_{{2}}(i)&f_{{1}}(j)+f_{{2}}(j)\end{matrix}\right)}$\foreignlanguage{english}{
and one has }
${0<Y_{{i,j}}^{{n}}<Y_{{i,i}}^{{n}},Y_{{j,j}}^{{n}}}$\foreignlanguage{english}{.
The distance matrix describes the X-tree in Fig. 6b. Two examples of
characters that are not independent are given in Fig.6c. }}

\bigskip

\includegraphics[width=12cm]{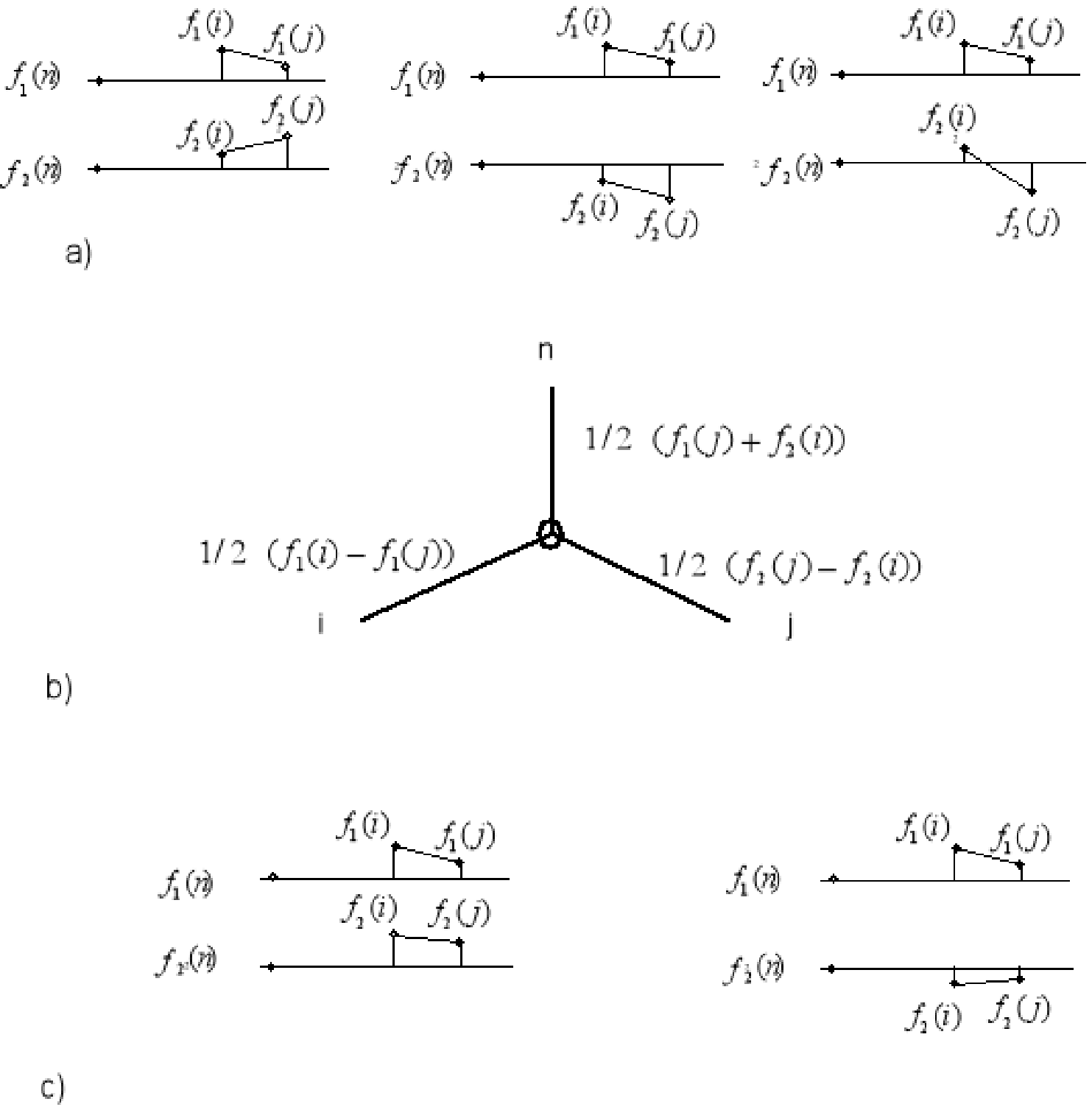}

\bigskip

\bigskip

{\selectlanguage{english}
\foreignlanguage{english}{Figure 6.}\foreignlanguage{english}{ a)
Examples of independent characters; b) X-tree corresponding to the
first two examples; c) The characters }
${f_{{1}}}$\foreignlanguage{english}{ and }
${f_{{2}}}$\foreignlanguage{english}{ are not independent.}}

\bigskip

{\selectlanguage{english}
\foreignlanguage{english}{Figure 7 is another illustration of
\ Proposition 3 for two characters on perfectly ordered taxa. The
ordered matrix }
${Y_{{i,j}}^{{n}}=Y_{{i,j}}^{{n}}(f_{{1}})+Y_{{i,j}}^{{n}}(f_{{2}})}$
\foreignlanguage{english}{is perfectly ordered. In this example, the
}\foreignlanguage{english}{distance matrix is described by a split
network and not by an X-tree (A tree is a special case among split
networks (Thuillard, 2007)).}}

 \includegraphics[width=12.351cm,height=3.967cm]{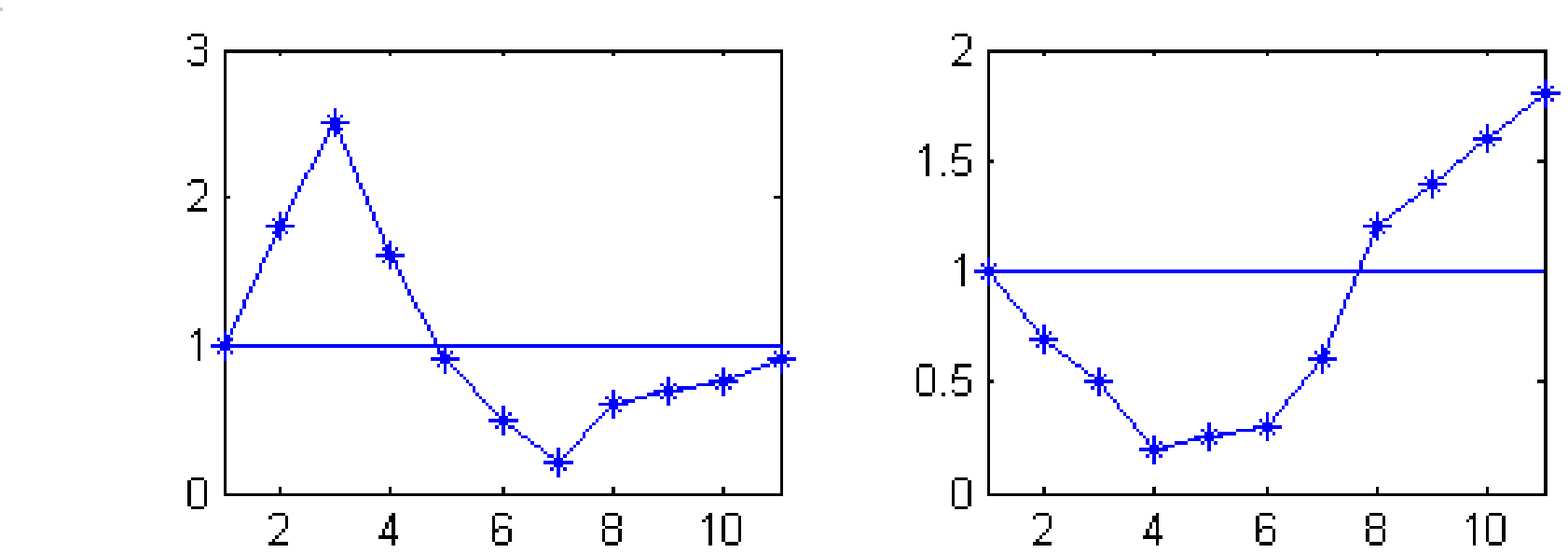} 

{\selectlanguage{english}
a)  \hskip 7 true cm       b)}

 \includegraphics[width=5.824cm,height=4.369cm]{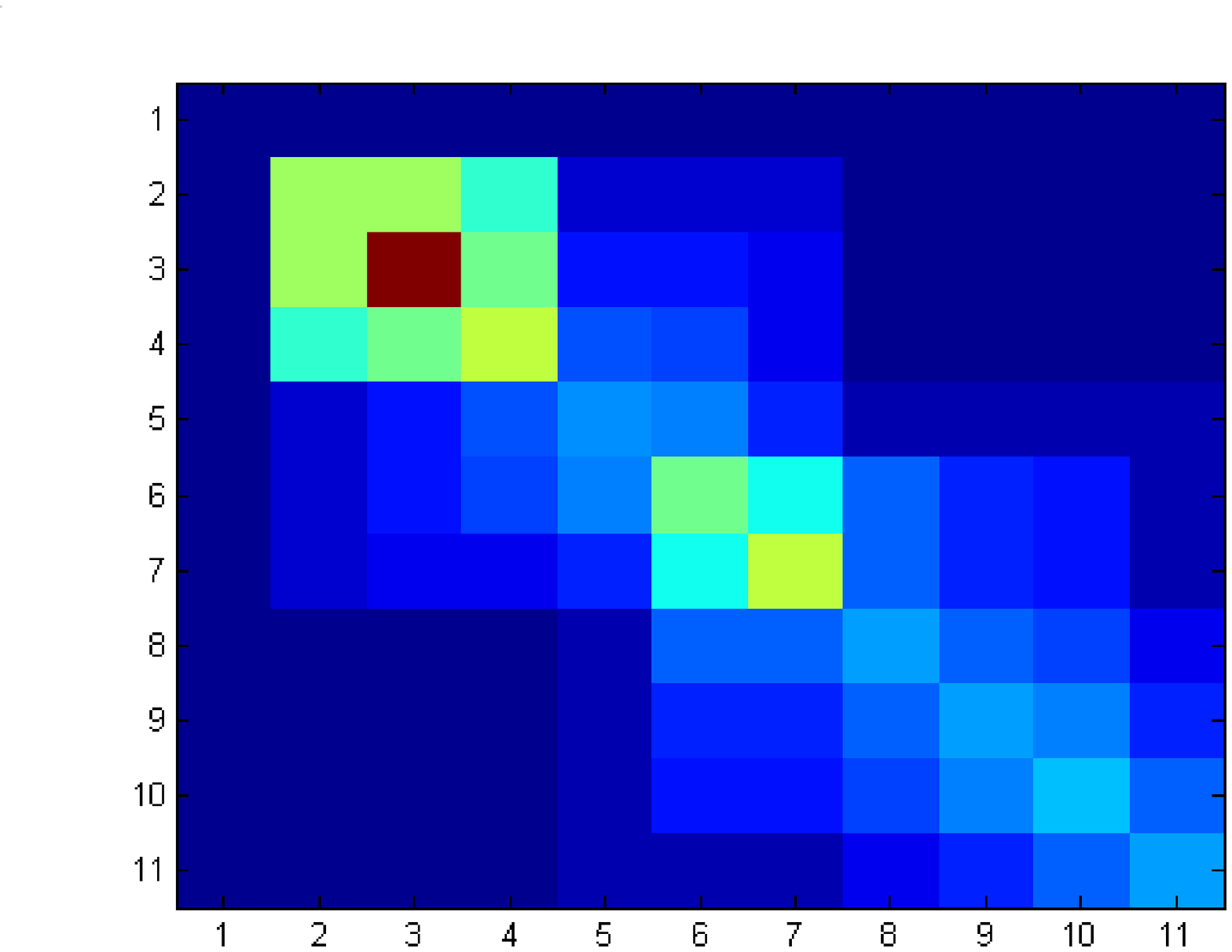} 
\includegraphics[width=0.369cm,height=3.741cm]{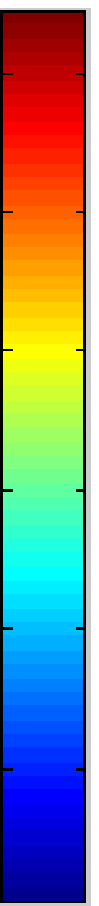}
\includegraphics[width=8.574cm,height=6.376cm]{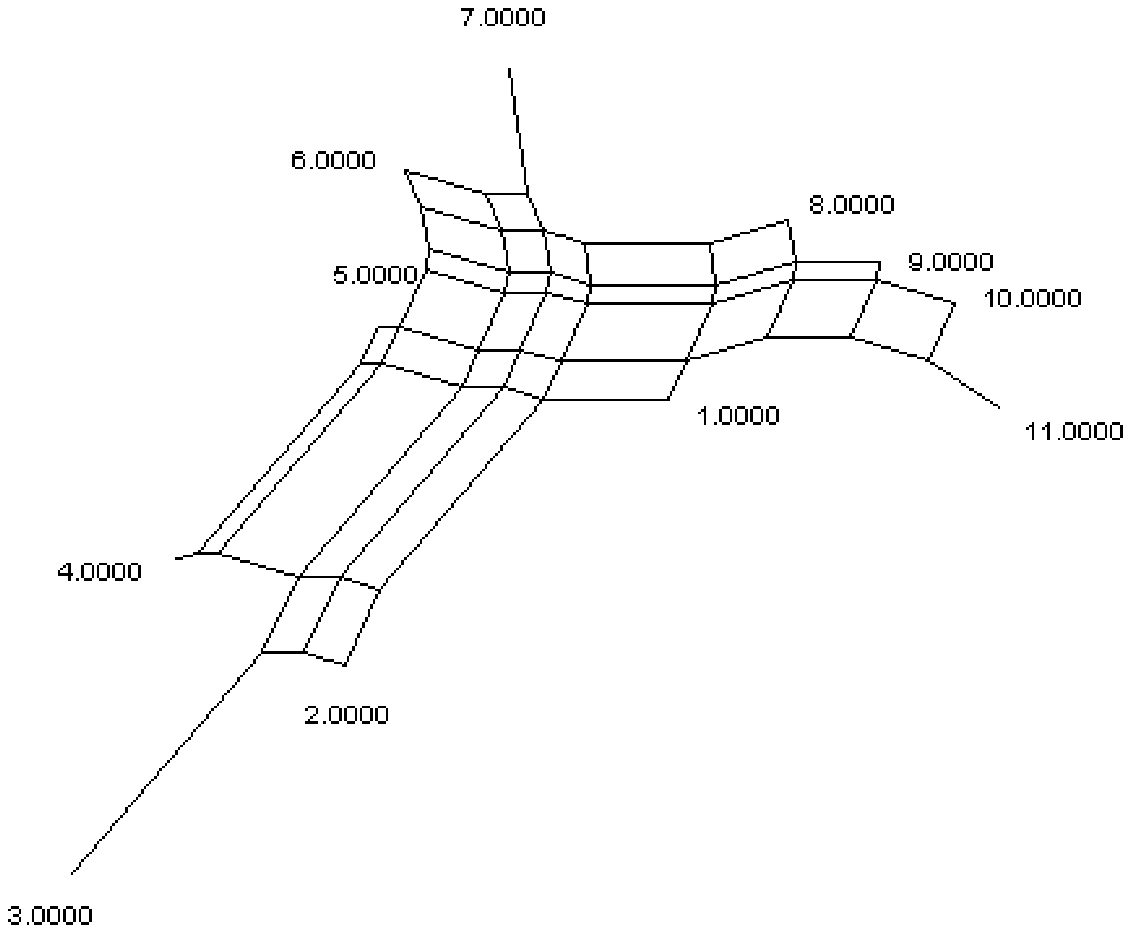} 

{\selectlanguage{english}
c)    \hskip 7 true cm   d)}

\bigskip

{\selectlanguage{english}
\foreignlanguage{english}{Figure 7.}\foreignlanguage{english}{ The
distance matrix } ${Y_{{i,j}}^{{n}}}$\foreignlanguage{english}{ (Fig.
7c) corresponding to two dependent characters } ${f_{{1}}(i)}$
\foreignlanguage{english}{and }
${f_{{2}}(i)}$\foreignlanguage{english}{. The distance matrix
corresponds to a split network (Fig. 7d). The split network is obtained
with Splits Tree (Huson and Bryant, 2006). The contradiction on the
order of the taxa is zero (C=0 in Eq. 2)}}

\bigskip

{\selectlanguage{english}\bfseries
5. Classification of Hominids Fossil Specimens}

{\selectlanguage{english}
\foreignlanguage{english}{The Minimum Contradiction on continuous
characters was tested on a set of independently analyzed data
representing craniofacial properties of hominid fossils. The results
obtained with the Minimum Contradiction Method are compared to
}\foreignlanguage{english}{those obtained with TNT in a recent article
in Nature. Gonz\'alez-Jos\'e et al. (2008) have analysed sets of
craniofacial landmarks representing the flexure of the cranial base,
facial retraction, neurocranial globularity, and masticatory apparatus.
Phylogenetic relationships among
}\foreignlanguage{english}{\textit{Homo}}\foreignlanguage{english}{
species and hominid taxa were obtained with the maximum parsimony
module for continuous characters in TNT. The reader is referred to
Gonz\'alez-Jos\'e et al. (2008) for the details on the extraction of
the data. \ }}

{\selectlanguage{english}
\foreignlanguage{english}{Similarly to Gonz\'alez-Jos\'e et al., we have
preprocessed the \ 4 sets of landmarks with the Generalized Procrustes
Analysis in Morphologika (O{\textquoteright} Higgins and Jones, 1998).
The Generalized Procrustes analysis is a superimposition method that
rotates, scales and translates the landmarks to adjust for isometric
effects of size and orientation. The distance between two taxa is
computed as the sum of the absolute difference between each Procrustes
coordinate. }\foreignlanguage{english}{The best circular order was
subsequently obtained by minimizing the contradiction C in Eq.(1)
(Thuillard, 2008). Figure 8 shows the minimum contradiction matrix
using }\foreignlanguage{english}{\textit{Gorilla
gorilla}}\foreignlanguage{english}{ as reference taxon.
}\foreignlanguage{english}{\textit{Gorilla
gorilla}}\foreignlanguage{english}{ is taken as the reference taxon in
order to be able to compare the results with Gonz\'alez-Jos\'e et al.
}}

{\selectlanguage{english}
\foreignlanguage{english}{The matrix } ${Y_{{i,j}}^{{n}}}$
\foreignlanguage{english}{is depicted using a colour coding. Large
values are coded red, while blue corresponds to small values of }
${Y_{{i,j}}^{{n}}}$\foreignlanguage{english}{. The minimum
contradiction matrix can be described as a split network. The order of
the taxa is quite compatible with the maximum parsimony tree of
Gonz\'alez-Jos\'e et al. A number of contradictions to perfect order
are observed for instance }\foreignlanguage{english}{\textit{H.
sapiens}}\foreignlanguage{english}{ vs
}\foreignlanguage{english}{\textit{H.
ergaster}}\foreignlanguage{english}{. As an example, let us describe
how the contradiction between }\foreignlanguage{english}{\textit{H.
sapiens}}\foreignlanguage{english}{ and
}\foreignlanguage{english}{\textit{H.
ergaster}}\foreignlanguage{english}{ can be extracted from Fig. 8. The
value \ } ${Y_{{9,\text{16}}}^{{n}}}$\foreignlanguage{english}{ is
coded in orange (45 on the right scale). The element }
${Y_{{9,\text{16}}}^{{n}}}$ \foreignlanguage{english}{is larger than
for instance }
${Y_{{9,\text{13}}}^{{n}}}$\foreignlanguage{english}{(Yellow=41) or }
${Y_{{\text{14},\text{16}}}^{{n}}}$\foreignlanguage{english}{=42. This
corresponds to a contradiction as according to the Kalmanson
inequalities, one should have } ${Y_{{9,\text{16}}}^{{n}}\le
Y_{{9,\text{13}}}^{{n}}}$\foreignlanguage{english}{ and }
${Y_{{9,\text{16}}}^{{n}}\le
Y_{{\text{14},\text{16}}}^{{n}}}$\foreignlanguage{english}{.
\ Contradictions in } ${Y_{{i,j}}^{{n}}}$\foreignlanguage{english}{
correspond to deviations from a tree or a split network structure
possibly caused by homoplasies or lateral transfers in genetic
sequences \ (Thuillard, 2008).}}

{\selectlanguage{english}
 \includegraphics[width=15.983cm,height=9.065cm]{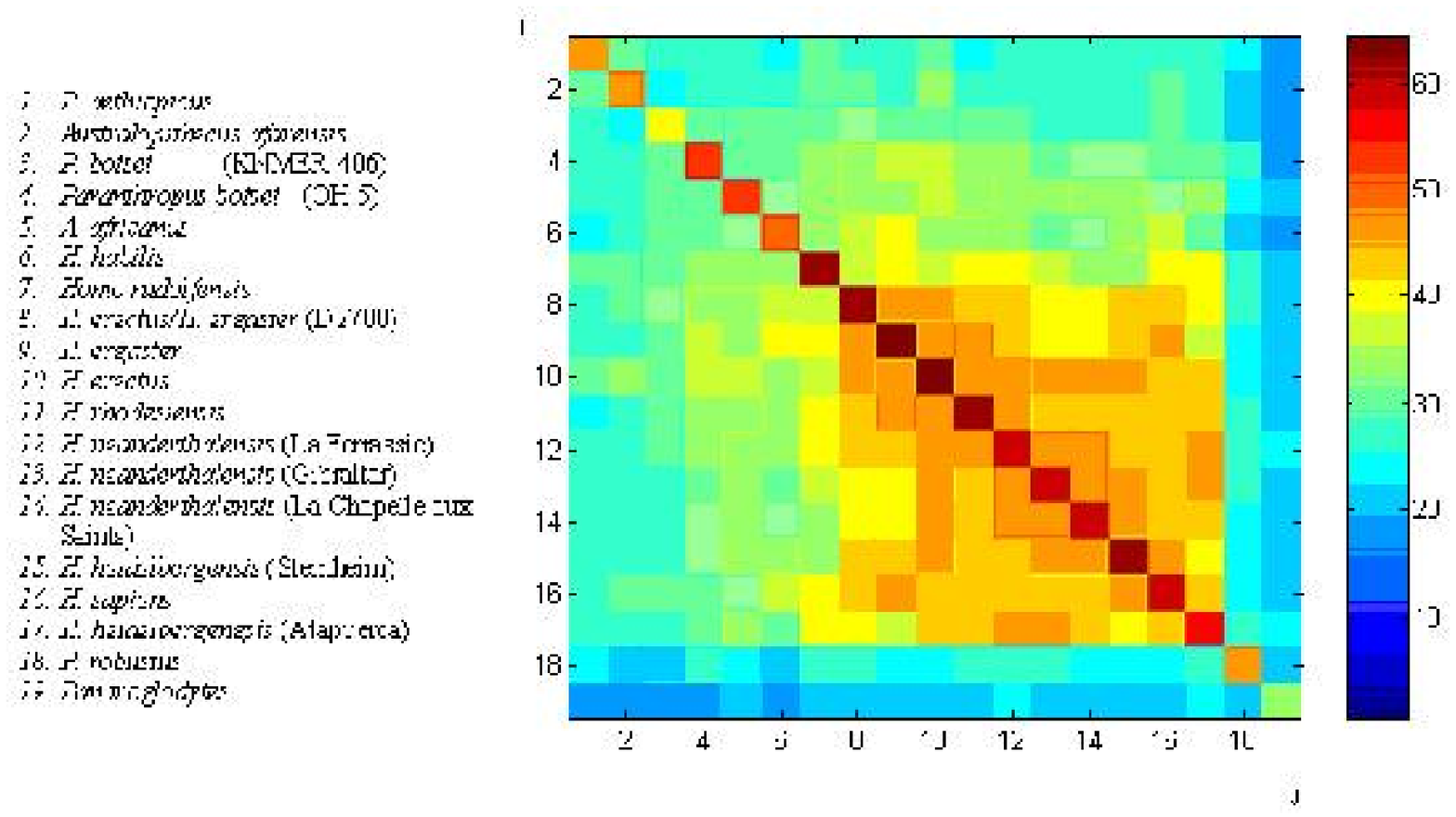}
\foreignlanguage{english}{Figure 8: Minimum contradiction matrix }
${Y_{{i,j}}^{{n}}}$ \foreignlanguage{english}{on a set of 20 hominid
taxa using }\foreignlanguage{english}{\textit{Gorilla
gorilla}}\foreignlanguage{english}{ as reference taxon n.}}

\bigskip

{\selectlanguage{english}
\foreignlanguage{english}{Table I shows the best order obtained with
\ the minimum contradiction approach and the order of the taxa on the
maximum parsimony tree. (The best order is a circular order and
}\foreignlanguage{english}{\textit{Gorilla
gorilla}}\foreignlanguage{english}{ is adjacent to both
}\foreignlanguage{english}{\textit{P.
aethiopicus}}\foreignlanguage{english}{ \ and
}\foreignlanguage{english}{\textit{Pan
troglodytes}}\foreignlanguage{english}{.) Except for
}\foreignlanguage{english}{\textit{H.
sapiens}}\foreignlanguage{english}{ the specimens are very similarly
ordered. The 2 main branches of the maximum parsimony tree \ are
indicated by a colour in the table. }}

\bigskip
\eject
{\selectlanguage{english}
\foreignlanguage{english}{Table I: Circular order obtained with the
Minimum Contradiction and the Maximum Parsimony approach on a set of
craniofacial landmarks of hominids (Maximum Parsimony order adapted
from }\foreignlanguage{english}{Gonz\'alez-Jos\'e et al. (2008)).}}

\begin{flushleft}
\tablehead{}\begin{tabular}{|m{7 cm}|m{7 cm}|}
\hline
\selectlanguage{english} \hfil Minimum Contradiction &
\selectlanguage{english} \hfil Maximum Parsimony\\\hline
\liststyleWWviiiNumii
\item \selectlanguage{english} 1. \itshape Gorilla gorilla
 &
\selectlanguage{english}\itshape Gorilla gorilla\\\hline
\liststyleWWviiiNumii
\setcounter{saveenum}{\value{enumi}}
\setcounter{enumi}{\value{saveenum}}
\item \selectlanguage{english}2. \itshape P. aethiopicus
 &
\selectlanguage{english}\itshape P. aethiopicus\\\hline
\liststyleWWviiiNumii
\setcounter{saveenum}{\value{enumi}}
\setcounter{enumi}{\value{saveenum}}
\item \selectlanguage{english}3. \itshape Australopithecus
afarensis
 &
\selectlanguage{english}\itshape Australopithecus afarensis\\\hline
\liststyleWWviiiNumii
\setcounter{saveenum}{\value{enumi}}
\setcounter{enumi}{\value{saveenum}}
\item \selectlanguage{english} \foreignlanguage{english}{4. \textit{P.
boisei  }}\foreignlanguage{english}{(KNMER-406)}
 &
\selectlanguage{english} \foreignlanguage{english}{\textit{P. boisei 
}}\foreignlanguage{english}{(KNMER-406)}\\\hline
\liststyleWWviiiNumii
\setcounter{saveenum}{\value{enumi}}
\setcounter{enumi}{\value{saveenum}}
\item \selectlanguage{english}
\foreignlanguage{english}{5. \textit{Paranthropus boisei 
}}\foreignlanguage{english}{(OH 5)}
 &
\selectlanguage{english} \foreignlanguage{english}{\textit{Paranthropus
boisei  }}\foreignlanguage{english}{(OH 5)}\\\hline
\liststyleWWviiiNumii
\setcounter{saveenum}{\value{enumi}}
\setcounter{enumi}{\value{saveenum}}
\item \selectlanguage{english}6. \itshape A. africanus
 &
\selectlanguage{english}\itshape A. africanus\\\hline
\liststyleWWviiiNumii
\setcounter{saveenum}{\value{enumi}}
\setcounter{enumi}{\value{saveenum}}
\item \selectlanguage{english}7. \itshape H. habilis
 &
\selectlanguage{english}\itshape H. habilis\\\hline
\liststyleWWviiiNumii
\setcounter{saveenum}{\value{enumi}}
\setcounter{enumi}{\value{saveenum}}
\item \selectlanguage{english}8. \itshape Homo rudolfensis
 &
\selectlanguage{english}\itshape Homo rudolfensis\\\hline
\liststyleWWviiiNumii
\setcounter{saveenum}{\value{enumi}}
\setcounter{enumi}{\value{saveenum}}
\item \selectlanguage{english} \foreignlanguage{ngerman}{9. \textit{H.
erectus/H. ergaster }}\foreignlanguage{ngerman}{(D2700)}
 &
\selectlanguage{english}
\foreignlanguage{ngerman}{\textit{\textcolor[rgb]{0.5019608,0.5019608,0.0}{H.
erectus/H. ergaster
}}}\foreignlanguage{ngerman}{\textcolor[rgb]{0.5019608,0.5019608,0.0}{(D2700)}}\\\hline
\liststyleWWviiiNumii
\setcounter{saveenum}{\value{enumi}}
\setcounter{enumi}{\value{saveenum}}
\item \selectlanguage{english}10. \itshape H. ergaster
 &
\selectlanguage{english}\itshape\color[rgb]{0.5019608,0.5019608,0.0} H.
ergaster\\\hline
\liststyleWWviiiNumii
\setcounter{saveenum}{\value{enumi}}
\setcounter{enumi}{\value{saveenum}}
\item \selectlanguage{english}11. \itshape H. erectus
 &
\selectlanguage{english}\itshape\color[rgb]{0.5019608,0.5019608,0.0} H.
erectus\\\hline
\liststyleWWviiiNumii
\setcounter{saveenum}{\value{enumi}}
\setcounter{enumi}{\value{saveenum}}
\item \selectlanguage{english}12. \itshape H. rhodesiensis
 &
\selectlanguage{english}\itshape\color[rgb]{0.5019608,0.5019608,0.0} H.
rhodesiensis\\\hline
\liststyleWWviiiNumii
\setcounter{saveenum}{\value{enumi}}
\setcounter{enumi}{\value{saveenum}}
\item \selectlanguage{english} \foreignlanguage{english}{13. \textit{H.
neanderthalensis }}\foreignlanguage{english}{(La
Ferrassie)}
 &
\selectlanguage{english}
\foreignlanguage{english}{\textit{\textcolor[rgb]{0.5019608,0.5019608,0.0}{H.}}}\foreignlanguage{english}{\textit{\textcolor[rgb]{0.5019608,0.5019608,0.0}{
sapiens}}}\\\hline
\liststyleWWviiiNumii
\setcounter{saveenum}{\value{enumi}}
\setcounter{enumi}{\value{saveenum}}
\item \selectlanguage{english} \foreignlanguage{english}{14. \textit{H.
neanderthalensis
}}\foreignlanguage{english}{(Gibraltar)}
 &
\selectlanguage{english}
\foreignlanguage{english}{\textit{\textcolor[rgb]{0.0,0.0,0.5019608}{H.
neanderthalensis
}}}\foreignlanguage{english}{\textcolor[rgb]{0.0,0.0,0.5019608}{(La
Ferrassie)}}\\\hline
\liststyleWWviiiNumii
\setcounter{saveenum}{\value{enumi}}
\setcounter{enumi}{\value{saveenum}}
\item \selectlanguage{english} \foreignlanguage{french}{15. \textit{H.
neanderthalensis}}\foreignlanguage{french}{ (La Chapelle aux
Saints)}
 &
\selectlanguage{english}
\foreignlanguage{french}{\textit{\textcolor[rgb]{0.0,0.0,0.5019608}{H.
neanderthalensis
}}}\foreignlanguage{french}{\textcolor[rgb]{0.0,0.0,0.5019608}{(La
Chapelle aux Saints)}}\\\hline
\liststyleWWviiiNumii
\setcounter{saveenum}{\value{enumi}}
\setcounter{enumi}{\value{saveenum}}
\item \selectlanguage{english} \foreignlanguage{english}{16. \textit{H.
heidelbergensis }}\foreignlanguage{english}{(Steinheim)}
 &
\selectlanguage{english}
\foreignlanguage{english}{\textit{\textcolor[rgb]{0.0,0.0,0.5019608}{H.
neanderthalensis
}}}\foreignlanguage{english}{\textcolor[rgb]{0.0,0.0,0.5019608}{(Gibraltar)}}\\\hline
\liststyleWWviiiNumii
\setcounter{saveenum}{\value{enumi}}
\setcounter{enumi}{\value{saveenum}}
\item \selectlanguage{english}17. \itshape H. sapiens
 &
\selectlanguage{english}
\foreignlanguage{english}{\textit{\textcolor[rgb]{0.0,0.0,0.5019608}{H.
heidelbergensis
}}}\foreignlanguage{english}{\textcolor[rgb]{0.0,0.0,0.5019608}{(Atapuerca)}}\\\hline
\liststyleWWviiiNumii
\setcounter{saveenum}{\value{enumi}}
\setcounter{enumi}{\value{saveenum}}
\item \selectlanguage{english} \foreignlanguage{english}{\textit{H.
heidelberg}}\foreignlanguage{english}{18. \textit{ensis
}}\foreignlanguage{english}{(Atapuerca)}
 &
\selectlanguage{english}
\foreignlanguage{english}{\textit{\textcolor[rgb]{0.0,0.0,0.5019608}{H.
heidelbergensis
}}}\foreignlanguage{english}{\textcolor[rgb]{0.0,0.0,0.5019608}{(Steinheim)}}\\\hline
\liststyleWWviiiNumii
\setcounter{saveenum}{\value{enumi}}
\setcounter{enumi}{\value{saveenum}}
\item \selectlanguage{english}19. \itshape P. robustus
 &
\selectlanguage{english}\itshape P. robustus\\\hline
\liststyleWWviiiNumii
\setcounter{saveenum}{\value{enumi}}
\setcounter{enumi}{\value{saveenum}}
\item \selectlanguage{english}20. \itshape Pan troglodytes
 &
\selectlanguage{english}\itshape Pan troglodytes\\\hline
\end{tabular}
\end{flushleft}
\color[rgb]{0.0,0.0,0.0}

\bigskip

{\selectlanguage{english}
\foreignlanguage{english}{Let us illustrate with an example the
possibilities offered \ by the Minimum Contradiction Method to analyze
phylogenetic data. In Fig.8, the largest values of }
${Y_{{i,j}}^{{n}}}$\foreignlanguage{english}{ for
i=}\foreignlanguage{english}{\textit{H.
habilis}}\foreignlanguage{english}{ and
}\foreignlanguage{english}{\textit{H.
rudolfensis}}\foreignlanguage{english}{ correspond to
j=}\foreignlanguage{english}{\textit{H.
ergaster}}\foreignlanguage{english}{ and
}\foreignlanguage{english}{\textit{H. sapiens
}}\foreignlanguage{english}{(}
${Y_{{i,j}}^{{n}}}$\foreignlanguage{english}{: yellow=41). Grouping
}\foreignlanguage{english}{\textit{H.
habilis}}\foreignlanguage{english}{ and
}\foreignlanguage{english}{\textit{H.
rudolfensis}}\foreignlanguage{english}{ with the other
}\foreignlanguage{english}{\textit{Homo}}\foreignlanguage{english}{
taxa is therefore a possibility. On the other hand }
${Y_{{i,j}}^{{n}}}$ \foreignlanguage{english}{has comparable values
within the cluster }\foreignlanguage{english}{\textit{H.
habilis}}\foreignlanguage{english}{,
}\foreignlanguage{english}{\textit{H. rudolfensis, A.
africanus}}\foreignlanguage{english}{,}\foreignlanguage{english}{\textit{
P. boisei}}\foreignlanguage{english}{ (KNMER-406), and
}\foreignlanguage{english}{\textit{Paranthropus
boisei}}\foreignlanguage{english}{ (OH 5). This offers a second
interpretation, namely that
}\foreignlanguage{english}{\textit{H.habilis}}\foreignlanguage{english}{
and }\foreignlanguage{english}{\textit{H.
rudolfensis}}\foreignlanguage{english}{ are related to
non}\foreignlanguage{english}{\textit{ Homo
}}\foreignlanguage{english}{taxa. In order to proceed with the
analysis, some definitions have to be introduced. Two consecutive taxa
with different character values define a cut. Two cuts in a circular
order define a split. A character is said to support a set of splits,
corresponding to all possible pairs of cuts, if after discretization of
the character{\textquoteright}s values the taxa are perfectly ordered.
(As a side remark, let us mention the connection existing between the
definition of a continuous character supporting a split and the
convexity of character states in a (non-valued) \ X-tree. If a
character supports a split on a valued X-tree then the \ character
states \ after discretization are convex (Semple and Steel, 2003)).}}

{\selectlanguage{english}
\foreignlanguage{english}{Contrarily to Gonz\'alez-Jos\'e et al., our
analysis is done without using a Principal Components Analysis (PCA).
This simplifies considerably the interpretation of the results.
Landmarks satisfying to a good approximation Prop. 1 can be identified
quite simply. Once those characters are identified, one can discover
which splits are supported by each character. Figure 9 shows a
character that supports the second interpretation of Fig. 8. The
landmark 9 (Facial retraction) supports a split between
}\foreignlanguage{english}{\textit{Homo}}\foreignlanguage{english}{
without }\foreignlanguage{english}{\textit{H.
habilis}}\foreignlanguage{english}{ and
}\foreignlanguage{english}{\textit{H.
rudolfensis}}\foreignlanguage{english}{ and
}\foreignlanguage{english}{the other taxa. In that example, both
interpretations are equally valid }(see also Cela-Conde and \ Amaya,
\ 2003)\foreignlanguage{english}{. }}

\bigskip

 \includegraphics[width=13cm]{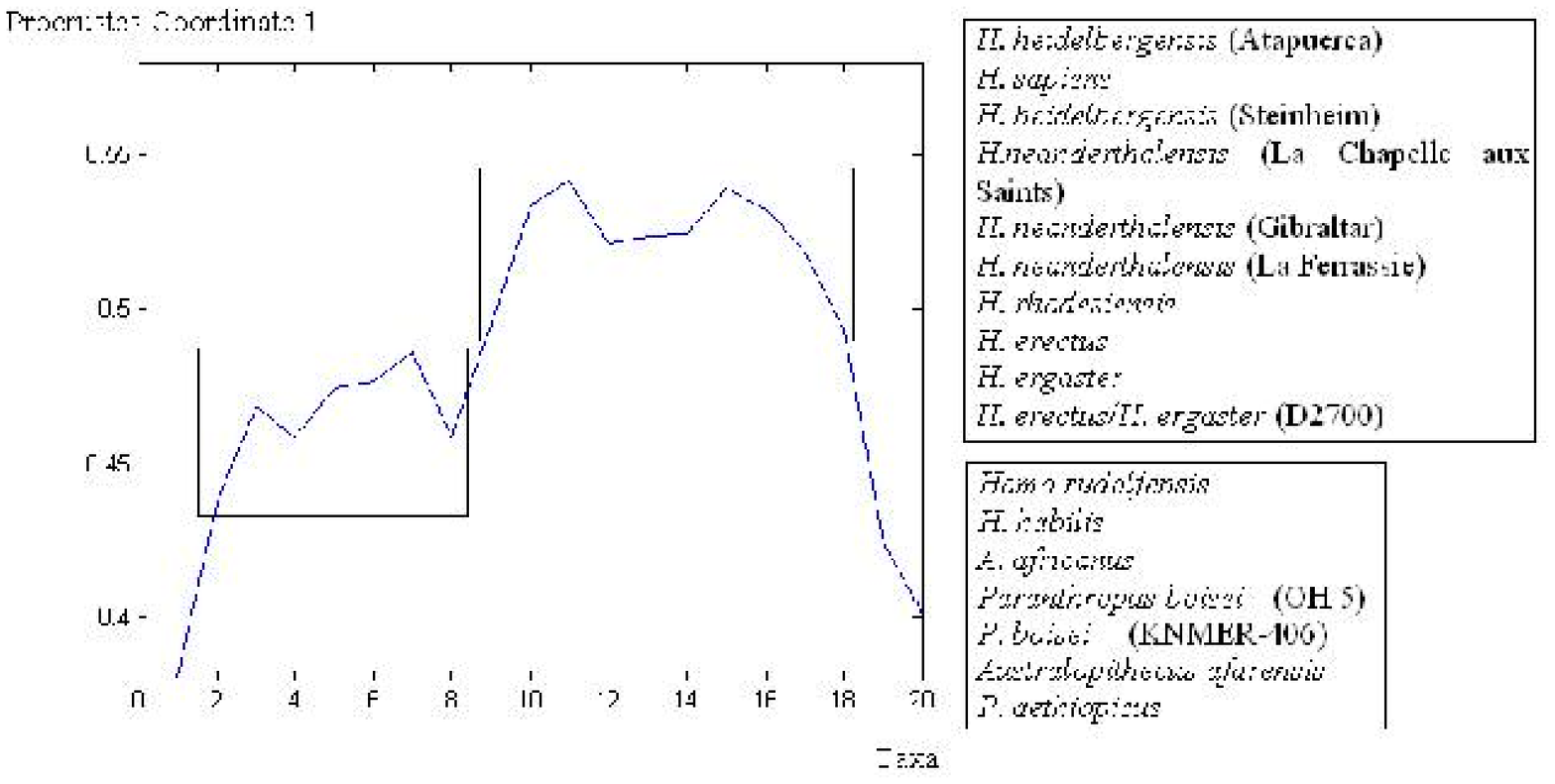} 

{\selectlanguage{english}
a)}

\bigskip

 \includegraphics[width=13cm]{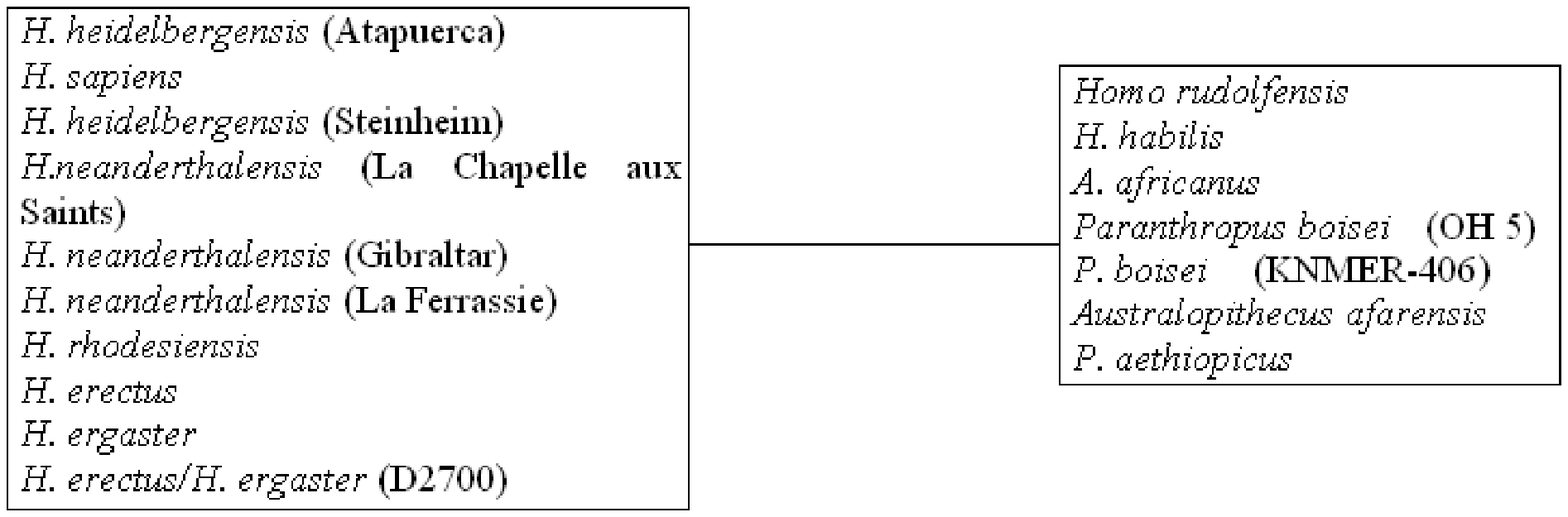} 

{\selectlanguage{english}
b)}

\bigskip

{\selectlanguage{english}
\foreignlanguage{english}{Figure 9: Examples showing how characters
supporting well a split can be identified using Prop. 1 in this
article. The order is the same as in Table I. a) The character
{\textquotedblleft}Facial retraction: landmark 9{\textquotedblright}
supports the split between
}\foreignlanguage{english}{\textit{Homo}}\foreignlanguage{english}{
without }\foreignlanguage{english}{\textit{H.
habilis}}\foreignlanguage{english}{ and
}\foreignlanguage{english}{\textit{H.
rudolfensis}}\foreignlanguage{english}{ and the other taxa. b) Split
for the character {\textquotedblleft}Facial retraction: landmark
9{\textquotedblright}.}}

\bigskip

{\selectlanguage{english}
\foreignlanguage{english}{The level of contradiction can be used as an
objective criterion to choose the reference node. As discussed in
details in Thuillard (2008,2009), the reference node is an important
choice in the presence of contradictions. In our example, the
normalized level of contradiction is lower if
}\foreignlanguage{english}{\textit{Pan troglodytes
}}\foreignlanguage{english}{is the reference taxon by about 30\%. This
suggests that }\foreignlanguage{english}{\textit{Pan
Troglodytes}}\foreignlanguage{english}{ is a better choice than
}\foreignlanguage{english}{\textit{Gorilla
Gorilla}}\foreignlanguage{english}{ as a reference taxon. Figure 10
shows quite interestingly that the ambiguity concerning
}\foreignlanguage{english}{\textit{H.
habilis}}\foreignlanguage{english}{ is removed with
}\foreignlanguage{english}{\textit{Pan troglodytes
}}\foreignlanguage{english}{as reference taxon.
}\foreignlanguage{english}{\textit{H.
habilis}}\foreignlanguage{english}{ belongs clearly to
}\foreignlanguage{english}{\textit{Homo}}\foreignlanguage{english}{. In
summary}\foreignlanguage{english}{\textit{,
}}\foreignlanguage{english}{with the data analyzed here,
}\foreignlanguage{english}{\textit{H.habilis}}\foreignlanguage{english}{
shares some characters with non
}\foreignlanguage{english}{\textit{Homo}}\foreignlanguage{english}{,
but has a majority of characters shared with other
}\foreignlanguage{english}{\textit{Homo}}\foreignlanguage{english}{
specimen, predominantly }\foreignlanguage{english}{\textit{H.erectus/H.
ergaster}}\foreignlanguage{english}{.}}

\bigskip

 \includegraphics[width=13cm]{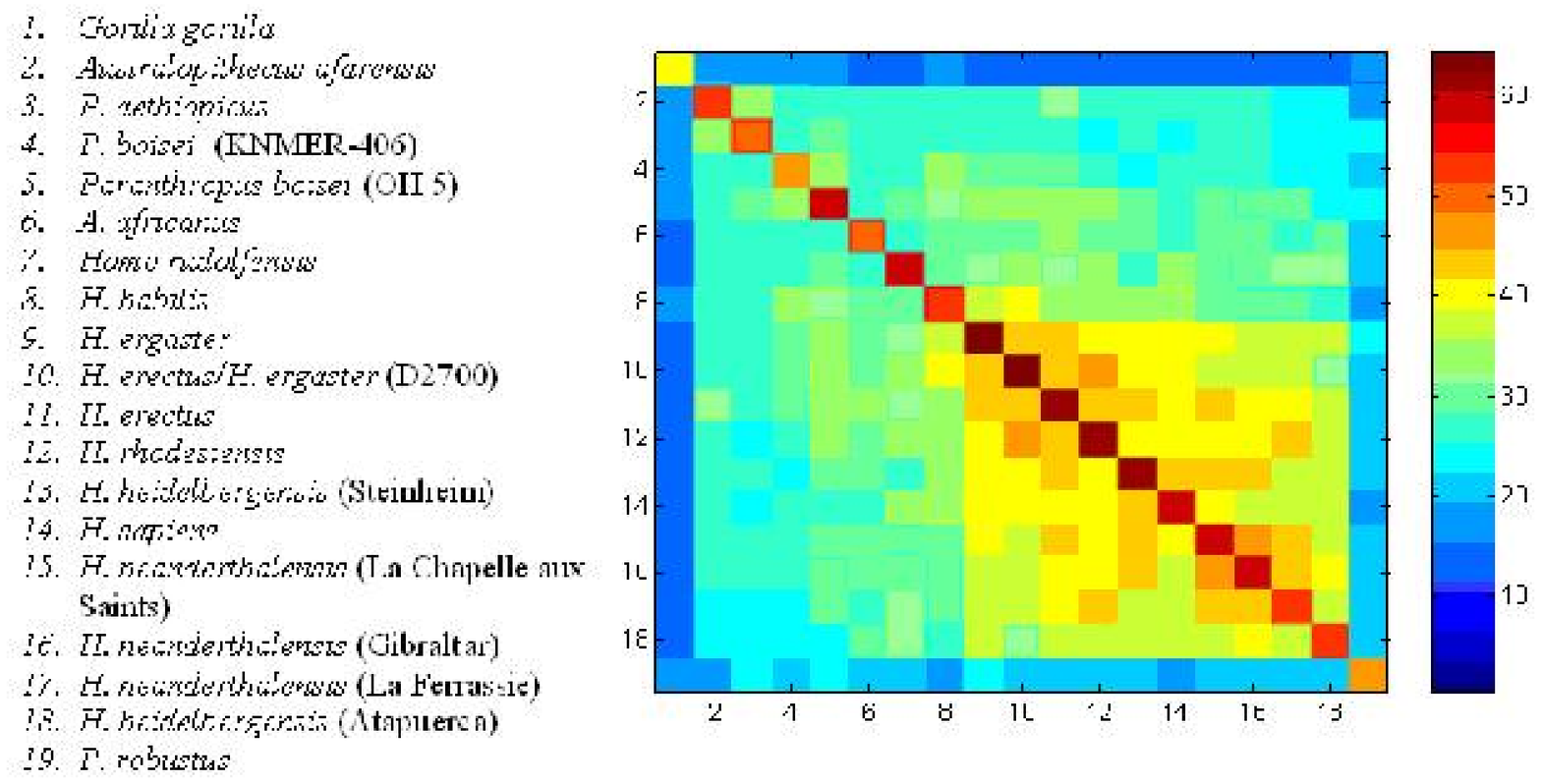} 

{\selectlanguage{english}
\foreignlanguage{english}{Figure 10: Minimum contradiction matrix }
${Y_{{i,j}}^{{n}}}$ \foreignlanguage{english}{on a set of 20 hominid
taxa using }\foreignlanguage{english}{\textit{Pan
Troglodytes}}\foreignlanguage{english}{ as reference taxon n.}}

\bigskip

{\selectlanguage{english}
A deeper analysis of the above results would go much beyond the goal of
this section. In this section we wanted to illustrate how information
can be extracted from a minimum contradiction analysis on continuous
variables. }

\bigskip

{\selectlanguage{english}
\foreignlanguage{english}{\textbf{6}}\foreignlanguage{english}{\textbf{.
Galaxies}}}

{\selectlanguage{english}
\foreignlanguage{english}{The second example}\foreignlanguage{english}{,
illustrating the continuous minimum contradiction approach, shows how a
character-based phylogenetic tree can be inferred from a distance
matrix. A standard approach to constructing phylogenetic trees from
continuous variables consists of discretizing the variables and to run
a maximum parsimony software treating the discretized variables as
characters. The difficulty with that approach is that the
discretization may easily disrupt an underlying tree structure. This
problem is particularly acute when 2-states characters are used. The
Minimum Contradiction Method can be applied to remedy that problem. Let
us explain the main idea on a 2-states character. Any perfectly ordered
variable f \ is transformed into a 2-states character C by the
following transformation: }
${C=\begin{matrix}1\overset{}{{}}f(i)>T\\0\overset{}{{}}f(i)\le
T\end{matrix}}$\foreignlanguage{english}{.}}

{\selectlanguage{english}
\foreignlanguage{english}{For illustration, we have taken from Ogando et
al (2008) a sample of 100 galaxies described by some observables and
derived quantities. In this section, our goal is to illustrate how the
Minimum \ Contradiction approach }\foreignlanguage{english}{can be used
in practice, in particular to discover structuring characters. The
astrophysical implications are out of the scope of the present work. It
will be presented in subsequent papers together with more in-depth
analysis. In practice, identifying a priori characters that behave like
on Figure 7a is difficult. For complex objects in evolution, this would
require some good knowledge of the evolution of the characters together
with some ideas about the correct phylogeny or at least a rough
evolutionary classification. In astrophysics, the study of galaxy
evolution has not yet reached this point (see e.g. Fraix-Burnet et al
2006a, 2006b, 2006c, 2009). However, we want to show here how the
approach presented in this paper can be extremely valuable even in
cases with very little a priori hints.}}

\bigskip

{\selectlanguage{english}
\foreignlanguage{english}{In this example, three variables are selected:
Brie, B-R, and OIII. Brie measures the surface brightness of the
galaxy, on a negative logarithm scale. B-R is the difference between
the B- and R-magnitudes: a high B-R indicates a red object (old stars
and/or high metallicity), while a low B-R indicates a blue object
(young stars and/or low metallicity). There is no a priori direct
physical connections between the three variables. High OIII (star
formation) could be expected to correspond to low B-R (young stars). As
shown in Fig. 11, that is not always true, due in large part to the
dependence of B-R on the metallicity of the stars.}}

\bigskip

 \includegraphics[width=6.503cm,height=4.89cm]{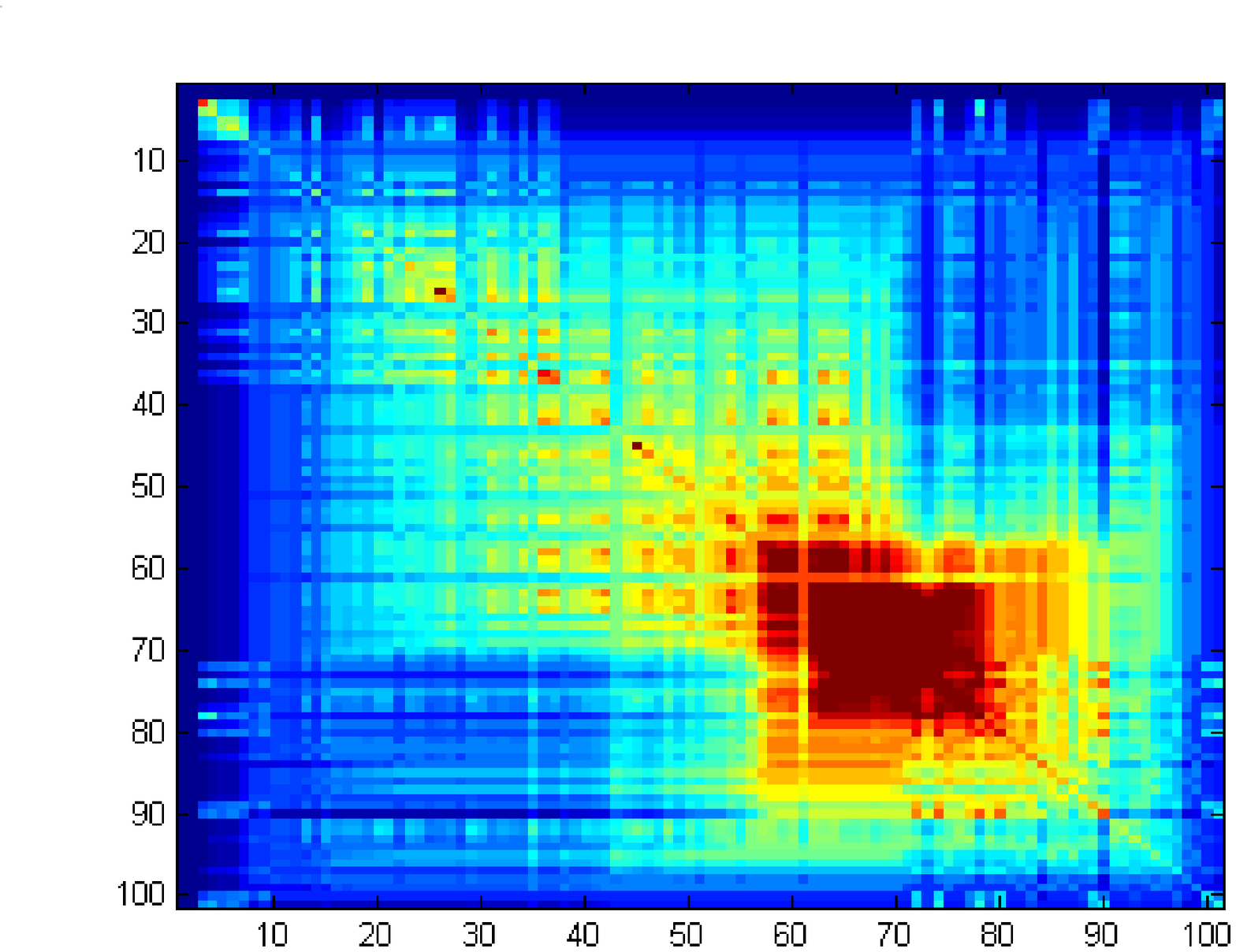} 
\includegraphics[width=0.582cm,height=4.972cm]{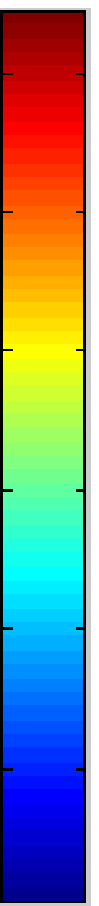}  
\includegraphics[width=7.174cm,height=8.662cm]{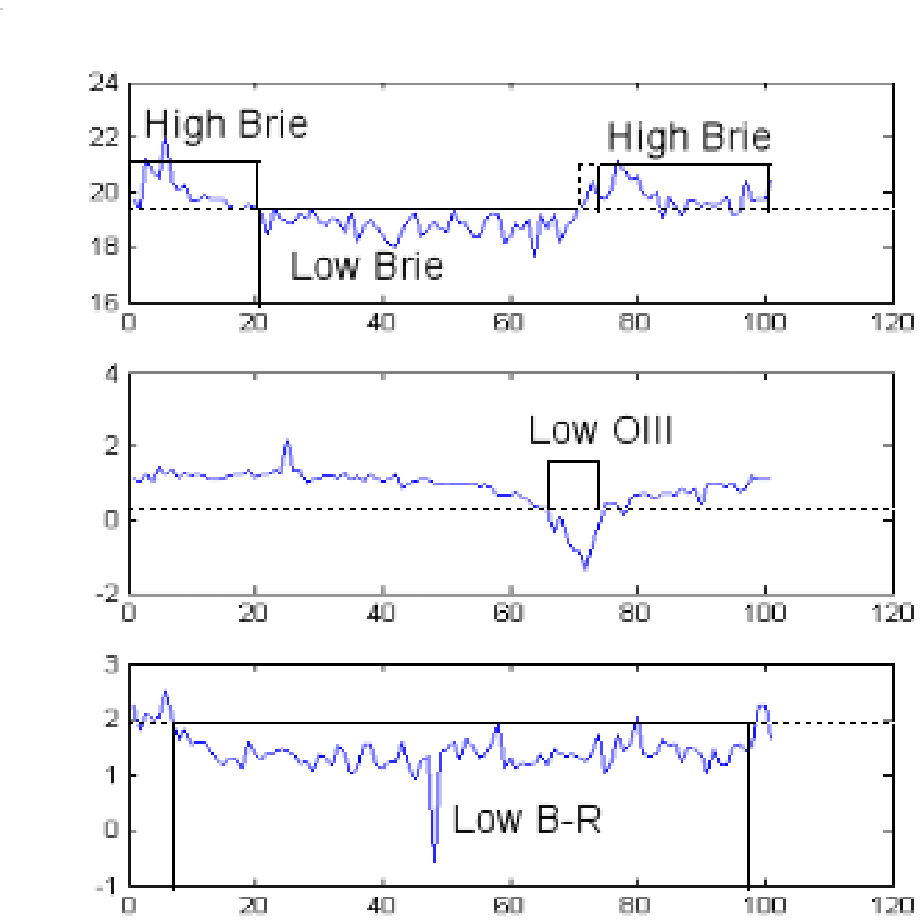}

\bigskip

{\selectlanguage{english}
a)     \hskip 7.5 true cm  b)}

\begin{center}
\includegraphics[width=10cm]{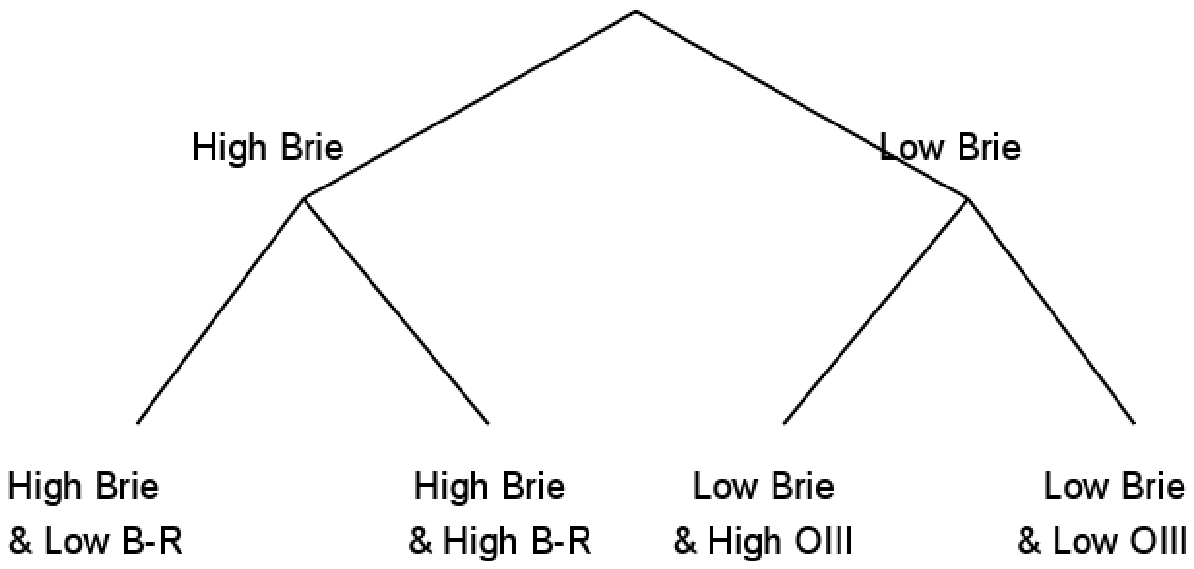}

\end{center}

{\selectlanguage{english}
c)}

\bigskip

{\selectlanguage{english}
\foreignlanguage{english}{Figure 11}\foreignlanguage{english}{. Analysis
of \ 3 selected characters Brie, OIII and B-R on an ensemble of 100
galaxies ordered with the Minimum Contradiction method. a) Distance
matrix }
${Y_{{\normalsubformula{\text{i,j}}}}^{{n}}}$\foreignlanguage{english}{;
b) Character values vs Galaxies after ordering: Top character Brie,
Middle: character OIII, Bottom: Character B-R; c) Tree describing
approximately the distance matrix after discretization (Solid line in
b).}}

\bigskip

{\selectlanguage{english}
\foreignlanguage{english}{After ordering, a number of clusters are
clearly recognized. The galaxies associated to the discrete character
{\textquotedblleft}High Brie{\textquotedblright} are far from being
perfectly ordered. The data }\foreignlanguage{english}{cannot be
described well with either a split network or a tree. This problem can
be solved by discretizing the variables. In Figure 11b, the 3 ordered
variables are represented together with a discretization of the input
variable using \ threshold values (dashed lines). Discretization
removes most contradictions on the order (In order to see it, let us
consider the character Brie. Let us code Brie High as 1 and Brie low as
0. The discretized function fulfils Prop. 1 as it has only a minimum
and any horizontal line crosses the discretized function at most
twice). The distance matrix corresponds well to a split network. The
split network can be represented, in first approximation, by an X-tree.
\ To do so let us move the boundary (dashed line) separating
{\textquotedblleft}low{\textquotedblright} from {\textquotedblleft}high
Brie{\textquotedblright} slightly to the right. The main split in the
tree corresponds to the {\textquotedblleft}High
Brie{\textquotedblright} and {\textquotedblleft}Low
Brie{\textquotedblright} branches. Each branch is split into two
\ other branches defined by the character states,
{\textquotedblleft}low OIII{\textquotedblright},
{\textquotedblright}High OIII{\textquotedblright} for
{\textquotedblleft}Low Brie{\textquotedblright} and
{\textquotedblleft}low B-R{\textquotedblright},
{\textquotedblleft}High-B-R{\textquotedblright} for
{\textquotedblleft}High-Brie{\textquotedblright}. The resulting tree is
\ shown in Figure 11b}}

{\selectlanguage{english}
\foreignlanguage{english}{The main splitting character is Brie for which
our discretization separates our sample in two roughly equal bins. That
is not the case for OIII and B-R for which low OIII and high B-R are
two small and distinct groups. All high Brie galaxies are in the high
OIII bin. Indeed, a low OIII corresponds to an absorption feature,
while a high OIII indicates an emission line due to star formation. As
a consequence, in this limited sample, low surface brightness galaxies
(main left branch) do ha}\foreignlanguage{english}{ve star formation,
and some high surface brightness objects show only an OIII absorption
feature (rightmost branch). All high B-R galaxies have high Brie and
high OIII. This means that in this sample, the red objects have a low
surface brightness, but they have some star formation. They are thus
not simply ageing galaxies, but probably form stars with high
metallicity. Conversely, all low OIII galaxies of our sample have a low
B-R, so that blue objects do not necessarily form a lot of stars. }}

{\selectlanguage{english}
\foreignlanguage{english}{A better understanding of the groupings and
their physical implications would require the investigation of other
properties of the objects. The relative complexity of the correlations
between our three characters implies that a correct classification
cannot be made by dichotomizing the variables beforehand. A more
objective and multivariate point of view is necessary to precise the
separating value between for instance
{\textquotedblleft}high{\textquotedblright} and
{\textquotedblleft}low{\textquotedblright} as in our present study.
Indeed, the discretization is here used only to depict more easily the
multivariate and continuous ordering of the objects in the sample. Fig.
11c is a synthetic classification shown by the distance matrix 11b and
obtained from the Minimum Contradiction method using fully continuous
information. }}

\bigskip

{\selectlanguage{english}
\foreignlanguage{english}{\textbf{7}}\foreignlanguage{english}{\textbf{.
Conclusions}}}

{\selectlanguage{english}
\foreignlanguage{english}{The }\foreignlanguage{english}{Minimum
Contradiction approach furnishes an objective justification to using
continuous variables or characters in phylogenetic studies. Provided
the taxa can be ordered so that each character fulfils the Kalmanson
inequalities then there exists a split network or a tree representing
exactly the distance matrix. We have shown that the Kalmanson
inequalities are fulfilled if the values of each character }can be
embedded into a function with at most a local maxima and a local
minima, and crossing any horizontal line at most twice.
\foreignlanguage{english}{In practical applications the level of
contradiction of the minimum contradiction matrix furnishes an
objective measure of the \ deviations to a tree or split network. This
approach was applied to a set of continuous characters, representing
faciocranial landmarks of hominids, already \ analyzed with a maximum
parsimony approach (Gonz\'alez et al., 2008). While the results are
found to be very similar to the maximum parsimony approach, the Minimum
Contradiction method furnishes supplementary information: i)
Problematic relationships between taxa \ are visualized. ii) Characters
supporting quite well a split can be discovered as they correspond to
single characters fulfilling very well the Kalmanson inequalities. iii)
Our approach can also }\foreignlanguage{english}{select the best
outgroup (reference taxon). The best outgroup leads to the order with
the smallest level of contradiction. }}

{\selectlanguage{english}
\foreignlanguage{english}{Discovering the structuring characters among a
set of continuous characters is a notoriously difficult task. The
search for structuring characters can be greatly facilitated by looking
for subsets of characters that satisfy best \ the Kalmanson
inequalities. This approach was applied to a set of 40 characters on
100 galaxies to extract the structuring characters. Quite
interestingly, while discretization of continuous characters is often
problematic, discretization \ with the Minimum Contradiction method can
help removing contradictions from a split network or tree structure. }}

\bigskip

{\selectlanguage{english}\sffamily\bfseries
Acknowledgements}

\bigskip

{\selectlanguage{english}
\foreignlanguage{english}{We thank Emmanuel Davoust for the compilation
of the data from the Ogando et al (2008) paper and from the Hyperleda
database
}\foreignlanguage{english}{(}\href{http://leda.univ-lyon1.fr/}{\textstyleInternetlink{\foreignlanguage{english}{http://leda.univ-lyon1.fr}}}\foreignlanguage{english}{).
Our thanks go also to Dr. R. Gonz\'alez-Jos\'e for his helpful
comments.}}

\bigskip

{\selectlanguage{english}\sffamily\bfseries
References}

\bigskip

{\selectlanguage{english}
\foreignlanguage{english}{Bandelt}\foreignlanguage{english}{, H.J. and
\ Dress, A. 1992. Split decomposition: a new and useful approach to 
phylogenetic analysis of distance data.
}\foreignlanguage{english}{\textit{Molecular Phylogenetic
Evolution}}\foreignlanguage{english}{ 1: 242-252.}}

{\selectlanguage{english}
\foreignlanguage{italian}{Cavalli-Sforza, L.L. and Edwards, A.W.F. 1967.
}\foreignlanguage{english}{Phylogenetic analysis: models and 
estimation procedures. }\foreignlanguage{english}{\textit{American
Journal of Human Genetics}}\foreignlanguage{english}{ 19:233-257.}}

{\selectlanguage{english}
\foreignlanguage{french}{Cela-Conde, C.J. and Ayala, F.J. 2003. }Genera
of the human lineage. \textit{Proc. Natl. Acad. Sci  USA }100:
7864-7869.}

{\selectlanguage{english}
\foreignlanguage{english}{Christopher}\foreignlanguage{english}{, G.E.,
Farach, M. and Trick, M.A. (1996) The structure of circular
decomposable  metrics. In European Symposium on Algorithms
(ESA){\textquoteright}96, Lectures Notes in  Computer Science 1136: pp
455-500.}}

{\selectlanguage{english}
\foreignlanguage{english}{Deineko, V., Rudolf, R. and Woeginger, G.
1995. Sometimes traveling is easy: the master  tour problem, Institute
of Mathematics, }\foreignlanguage{english}{\textit{SIAM Journal on
Discrete Mathematics}}\foreignlanguage{english}{ 11: 81  {}- 93.}}

{\selectlanguage{english}
\textstyleHTMLZitat{\foreignlanguage{ngerman}{\textup{Eisen,}}}\textstyleHTMLZitat{\foreignlanguage{ngerman}{\textup{
M.B, Spellman, P.T., Brown, P.O. and Botstein, D. 1998.
}}}\textstyleHTMLZitat{\textup{Cluster analysis and display  of 
genome-wide expression patterns.
}}\foreignlanguage{italian}{\textit{Proc. Natl. Acad.
Sci.}}\textstyleHTMLZitat{\foreignlanguage{italian}{\textup{
}}}\textstyleHTMLZitat{\foreignlanguage{italian}{USA}}\textstyleHTMLZitat{\foreignlanguage{italian}{\textup{
95: 14863-- 14868.}}}}

{\selectlanguage{english}
\foreignlanguage{italian}{Edwards, A.W.F. and Cavalli-Sforza, L.L. 1964.
}\foreignlanguage{english}{Reconstruction of evolutionary trees. pp.
67- 76. \ In Phenetic and Phylogenetic Classification, ed. V. H.
Heywood and J. McNeill.  Systematics Association pub. no. 6, London.}}

{\selectlanguage{english}
Felsenstein, J. 2004. Inferring phylogenies, Sinauer Associates.}

{\selectlanguage{english}
\foreignlanguage{english}{Fraix-Burnet, D. Choler, P., Douzery, E.,
Verhamme, A. 2006a Astrocladistics: a }\foreignlanguage{english}{
phylogenetic analysis of galaxy evolution. I. Character evolutions and
galaxy histories.  }\foreignlanguage{english}{\textit{Journal of
Classification}}\foreignlanguage{english}{ 23, 31-56.
(}\url{http://arxiv.org/abs/astro-ph/0602581}\foreignlanguage{english}{).}}

{\selectlanguage{english}
\foreignlanguage{english}{Fraix-Burnet, D., Douzery, E., Choler, P.,
Verhamme, A. 2006b. Astrocladistics: a }\foreignlanguage{english}{
phylogenetic analysis of galaxy evolution. II. Formation and
diversification of  galaxies.
}\foreignlanguage{english}{\textit{Journal of
Classification}}\foreignlanguage{english}{ 23, 57-78.
(http://arxiv.org/abs/astro-ph/0602580)}}

{\selectlanguage{english}
\foreignlanguage{french}{Fraix-Burnet, D., Choler, P., Douzery, E.
2006c. }\foreignlanguage{english}{Towards a phylogenetic analysis of
galaxy  evolution: a case study with the dwarf galaxies of the local
group. Astronomy \&  Astrophysics 455, 845-851.
(http://arxiv.org/abs/astro-ph/0605221).}}

{\selectlanguage{english}
\foreignlanguage{english}{Fraix-Burnet, D. 2009. Galaxies and
Cladistics. In Evolutionary Biology from Concept to  Application II,
Springer, in press.}}

{\selectlanguage{english}
Goloboff, P., Farris, J. and Nixon, K. 2008. TNT: a free program for
phylogenetic }

{\selectlanguage{english}
\foreignlanguage{english}{ }\foreignlanguage{english}{analysis.
}\foreignlanguage{english}{\textit{Cladistics}}\foreignlanguage{english}{
24: 774-786.}}

{\selectlanguage{english}
\foreignlanguage{english}{Gonz\'alez-Jos\'e, R., Escapa, I., Neves,
W.A., H\'ector, R.C., Pucciarelli, M. 2008. Cladistic
}\foreignlanguage{english}{ analysis of continuous modularized traits
provides phylogenetic signals in Homo  evolution.
}\foreignlanguage{english}{\textit{Nature}}\foreignlanguage{english}{
453: 775-778.}}

{\selectlanguage{english}
\foreignlanguage{english}{Huson, D. and Bryant, D. 2006 Application of
phylogenetic networks in evolutionary studies.
}\foreignlanguage{english}{ }\foreignlanguage{english}{\textit{Mol.
Biol. Evol}}\foreignlanguage{english}{.23(2):254-267.}}

{\selectlanguage{english}
\foreignlanguage{english}{Kalmanson, K. 1975. Edgeconvex circuits and
the traveling salesman problem.
}\foreignlanguage{english}{\textit{Canadian  Journal of
Mathematics}}\foreignlanguage{english}{ 27: 1000-1010.}}

{\selectlanguage{english}
\foreignlanguage{english}{Kunin V, Ahren D, Goldovsky L, Janssen P and
Ouzounis CA. 2005}\foreignlanguage{english}{. Measuring genome 
conservation across taxa: divided strains and united kingdoms. Nucleic
Acids  Research, 33(2): 616-621.}}

{\selectlanguage{english}
\foreignlanguage{english}{Lee, C., Blay, S.,
Mooers,}\foreignlanguage{english}{ A.O., Singh, A., and Oakley, T.H.
2006. CoMET: A Mesquite  package for comparing models of continuous
character evolution on phylogenies.  }\textit{Evolutionary
Bioinformatics }2: 183-186.}

{\selectlanguage{english}
\foreignlanguage{english}{MacLeod, N. and Forey, P.L. 2003. Morphology,
Shape and Phylogeny, Eds. Taylor and }\foreignlanguage{english}{
Francis Inc., New York.}}

{\selectlanguage{english}
\foreignlanguage{english}{Makarenkov}\foreignlanguage{english}{, V. and
Leclerc, B. 1997. Circular orders of tree metrics, and their uses for
the  reconstruction and fitting of phylogenetic trees. In Mirkin, B.,
Morris F.R., Roberts,  F., Rzhetsky, A, eds. Mathematical hierarchies
and Biology, DIMACS Series in  Discrete Mathematics and Theoretical
Computer Science. Providence: Amer. Math.  Soc. pp 183-208.}}

{\selectlanguage{english}
\foreignlanguage{italian}{Ogando, R.L.C., Maia, M.A.G., Pellegrini,
P.S., da Costa, L.N. 2008. }\foreignlanguage{english}{\textit{The
Astronomical  Journal}}\foreignlanguage{english}{, 135, 2424-2445
(}\url{http://fr.arxiv.org/abs/0803.3477}\foreignlanguage{english}{).}}

{\selectlanguage{english}
\foreignlanguage{english}{Oakley, T.H. and Cunningham C.W. 2000.
Independent contrasts succeed where ancestor
}\foreignlanguage{english}{ reconstruction fails in a known
bacteriophage phylogeny.
}\foreignlanguage{english}{\textit{Evolution}}\foreignlanguage{english}{
54 (2), 397-405.}}

{\selectlanguage{english}
\foreignlanguage{english}{O{\textquoteright}
Higgins}\foreignlanguage{english}{, P. and Jones, N. 1998. Facial
growth in Cercocebus torquatus: An application of  three  dimensional
geometric morphometric techniques to the study of  morphological~
variation. }\foreignlanguage{english}{\textit{Journal of
Anatomy}}\foreignlanguage{english}{ 193: 251-272.}}

{\selectlanguage{english}
\foreignlanguage{english}{Planet, P}\foreignlanguage{english}{.J,
DeSalle, R., Siddal, M., Bael, T., Sarkar, I.N., Stanley, S.E. 2001.
Systematic  analysis of DNA microarray data: ordering and interpreting
patterns of gene  expression .
}\foreignlanguage{english}{\textit{Genome
Research}}\foreignlanguage{english}{ 11: 1149-1155.}}

{\selectlanguage{english}
Semple, C. and Steel, M. 2003. Phylogenetics, Oxford University Press,
New York.}

{\selectlanguage{english}
\foreignlanguage{english}{Thuillard}\foreignlanguage{english}{, M. 2007.
Minimizing contradictions on circular order of phylogenic trees. 
}\foreignlanguage{english}{\textit{Evolutionary
Bioinformatics}}\foreignlanguage{english}{ 3: 267-277.}}

{\selectlanguage{english}
\foreignlanguage{english}{Thuillard, M.}\foreignlanguage{english}{2008.
\ Minimum contradiction matrices in whole genome phylogenies. 
}\foreignlanguage{english}{\textit{Evolutionary
Bioinformatics}}\foreignlanguage{english}{ 4: 237-247.}}

{\selectlanguage{english}
\foreignlanguage{english}{Thuillard}\foreignlanguage{english}{, M. 2009.
Why phylogenetic trees are often quite robust against lateral
transfers. In  Evolutionary Biology from Concept to Application II,
Springer, in press.}}

\bigskip

\bigskip

\bigskip

\bigskip

\bigskip

\bigskip

\bigskip

\bigskip

\bigskip

\bigskip

\bigskip

\bigskip

\bigskip

\bigskip

\bigskip

\bigskip

\bigskip

\bigskip

\bigskip

\bigskip

\bigskip

\bigskip

\bigskip

\bigskip

\bigskip

\bigskip

\bigskip

\bigskip

\bigskip

\bigskip

\bigskip

\bigskip

\bigskip

\bigskip
\end{document}